\documentclass[preprintnumbers,eqsecnum,aps,prd,epsf,showpacs,nofootinbib, 
twocolumn]{revtex4}
\usepackage{latexsym,amsmath,amssymb}
\usepackage[dvips]{graphicx,color}% Include figure files
\usepackage{dcolumn,bm}    % Align table columns on decimal point
\usepackage{subfigure}  % sub labeling of figure
\usepackage{color}
\input{colordvi.tex}
\newcommand{\gsim}{\mbox{\raisebox{-1.ex}{$\stackrel
      {\textstyle>}{\textstyle\sim}$}}}

\def\til#1{\tilde{#1}} 
\newcommand{\calO}{{\cal O}}

\begin{document}
\thispagestyle{empty}
\title{Bubble Universes With 
Different Gravitational Constants}

\author{Yu-ichi Takamizu$^{1}$}
\email{yt313@cam.ac.uk}
\author{Kei-ichi Maeda$^{2}$}
\email{maeda@waseda.jp}

\affiliation{$^{1}$ 
Department of Applied Mathematics and Theoretical Physics, University of Cambridge, Wilberforce Road, Cambridge CB3 0WA, UK\\
$^{2}$ Department of Physics, Waseda University,
Okubo 3-4-1, Shinjuku, Tokyo 169-8555, Japan}
\date{\today}

\begin{abstract}
We argue a scenario motivated by the context of string landscape, 
where our universe is produced by a new vacuum bubble embedded in an old 
bubble and these bubble universes have not only 
different cosmological constants, but also their own different gravitational constants. We study these effects on the primordial curvature perturbations. In order to construct a model of varying gravitational constants, 
we use the Jordan-Brans-Dicke (JBD) theory where different 
expectation values of scalar fields produce difference of constants. In this system, we investigate the nucleation of bubble universe and dynamics of the wall separating two spacetimes. 
In particular, the primordial curvature perturbation on superhorizon scales can be affected by the wall trajectory as the boundary effect. 
We show the effect of gravitational constant in the exterior bubble universe can provide a peak like a bump feature at a large scale 
in a modulation of power spectrum. 
\end{abstract}
\maketitle
%%%%%%%%%%%%%%%%%%%%%%%%%%%%%%%%%%%%%%%%%%%%%%%%%%%%%%%%
\section{introduction}
In the context of the string theory landscape \cite{Susskind:2003kw}, there are various 
possibilities that multiple vacua can be 
produced whose semiclassical tunnelling process from 
metastable vacua to new vacua can occur. 
In this respect, it can allow us to argue that 
nucleation of spherical symmetric regions, namely Bubble in space which 
are constructed by a new vacuum and expand into the old vacuum. 
One of any bubble universe 
created by such first-order phase transition may be 
regarded as our universe \cite{Basu:1991ig,Coleman:1980aw,Sato:1980yn,Sato:1981gv,Maeda:1985ye,Berezin:1987bc,Bucher:1994gb,Yamamoto:1995sw,Garriga:1998he,Aguirre:2007an,Simon:2009nb,Casadio:2011jt,KeskiVakkuri:1996gn}. 
In particular, inspired by the 
bubble universe scenario, an accelerated expansion of the universe 
generated by its vacuum energy would give a natural explanation for inflation realization in the early universe as discussed old inflation \cite{Sato:1980yn} 
or open inflation \cite{Bucher:1994gb,Yamamoto:1995sw,Garriga:1998he}
scenario in the past. 
If the bubble universe would be inside the old bubble 
universe, the so-called parent universe, we are mainly interested in 
how much information we know or detect through any observational signature 
of the parent universe.  

In the viewpoint of Anthropic principle \cite{Susskind:2003kw,Weinberg:1988cp}, it is possible that several universes take their own choices of natural constants. It can allow us to 
argue a possibility that landscape vacua have different values of gravitational constants as well as cosmological constants.  In this paper, the bubble universes have their own different gravitational constants and it can give us explore 
how a different gravitational constant in the 
exterior universe affect and relates to 
physical (especially observational) quantities 
for the observer in the interior bubble universe 
(see also works \cite{Aguirre:2007an} 
for the aim). 

In order to such model building that each bubble universe has its own 
different gravitational constant, 
it is useful to employ the JBD theory \cite{Brans:1961sx} 
(see book \cite{Fujii:2003pa} for review). This theory can 
achieve a varying gravitational constant by identified 
a vacuum expectation value of the scalar field with gravitational constants in the original, the so-called Jordan frame. Therefore we will use this theory to construct our model and investigate the consequence of the model to our (child) 
bubble universe, where the child bubble is inside the parent bubble with different constants. 

Inflationary scenario  
can give a remarkably successful way of generating a consistent spectrum of 
curvature perturbations with recent cosmological observations, such as the cosmic microwave background (CMB) and galaxy survey to explore a 
large-scale structure \cite{Sato:1980yn,Guth:1980zm,Lyth-book}. 
The quantum fluctuations generated in a exponential expansion of the early stage of the universe 
were spread out a cosmological horizon scale and their wavelengths are then 
pushed to superhorizon size, which can be reduced to classical perturbations 
seeding cosmic structure. After inflation ends, they reenter the horizon in a radiation dominated phase and can be detected by such as CMB observation (see \cite{Lyth-book} for review). 
The recent data by Planck satellite \cite{Ade:2013ktc} gave a perfectly agreement with the prediction of inflationary theory. 

In order to discuss an observational consequence of bubble universes 
with different gravitational constants, we analyze density perturbations
during an inflationary phase based on the JBD theory.
Note that although several inflationary models with/without bubble 
formations and the accompanying density perturbations are discussed based on 
the scalar-tensor gravity theories including the JBD  theory
\cite{extended_inflation,soft_inflation,monopole_inflation}, 
the JBD scalar field is assumed to be an inflaton in those models.
In the present paper, however, we assume that there exists an inflaton field 
in addition to the JBD scalar field, which is used to
fix only two different gravitational constants. 

If the superhorizon perturbations propagate even 
though their amplitudes freeze out and approach to the boundary of the bubble universe, they would be scattered by the layer called bubble wall, separating a new and old vacuum. Namely when one join the interior and exterior perturbation mode over the bubble wall, a reflected wave can be produced. Hence if there would be a possibility  that  one can 
get information of outside bubble,  that is only way to analyze perturbations 
coming back from superhorizon scales outside bubble as a trace of the 
outer region (see \cite{Ertan:2007eq} in a different setup, but 
in which a similar effect has been studied).  

The aim of this paper is to obtain physical signature on a primordial 
curvature perturbation at superhorizon scales, which is 
affected by the boundary of bubble. 
In particular, we can provide how superhorizon mode is modified by the effect of different gravitational constant in the exterior universe by employing the JBD  theory. 

The paper is organized as follows. In section 2 we show our model setup and 
discuss a nucleation of bubble universe and a static bubble wall solution. 
In section 3, we study dynamics of bubble wall after a nucleation. 
In section 4, we adopt linear perturbation theory to the system and match perturbation modes in both 
sides over the bubble wall. In particular, our calculation is the case for 
de Sitter background. Then we discuss the effect of different gravitational constant on the obtained results in section 5. Section 6 is devoted to conclusion. 
%%%%%%%%%%%%%%%%%%%%%%%%%%%%%%%%%%%%%%%
\section{model}
We consider the model of bubble universes which have 
their own gravitational constants. The wall of the bubble can divide two bubble universes whose constants have 
different values. We investigate the effect of different gravitational constant on the superhorizon scale perturbation. 

\begin{figure*}[t]
\vspace{-1cm}
\begin{center}
\begin{tabular}{lll}
\includegraphics[width=5cm]{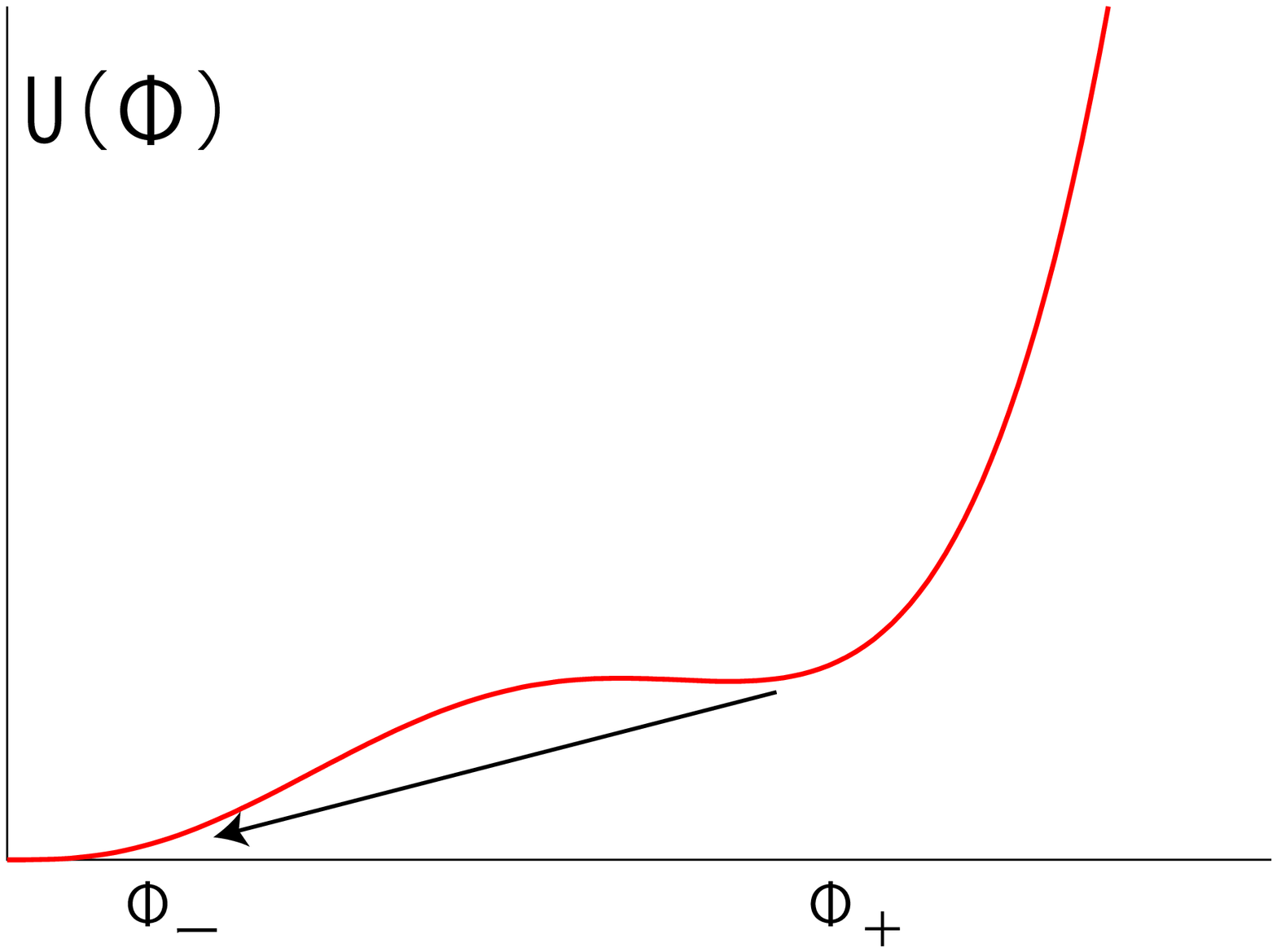} & 
\hspace{0.5cm}
\includegraphics[width=5cm]{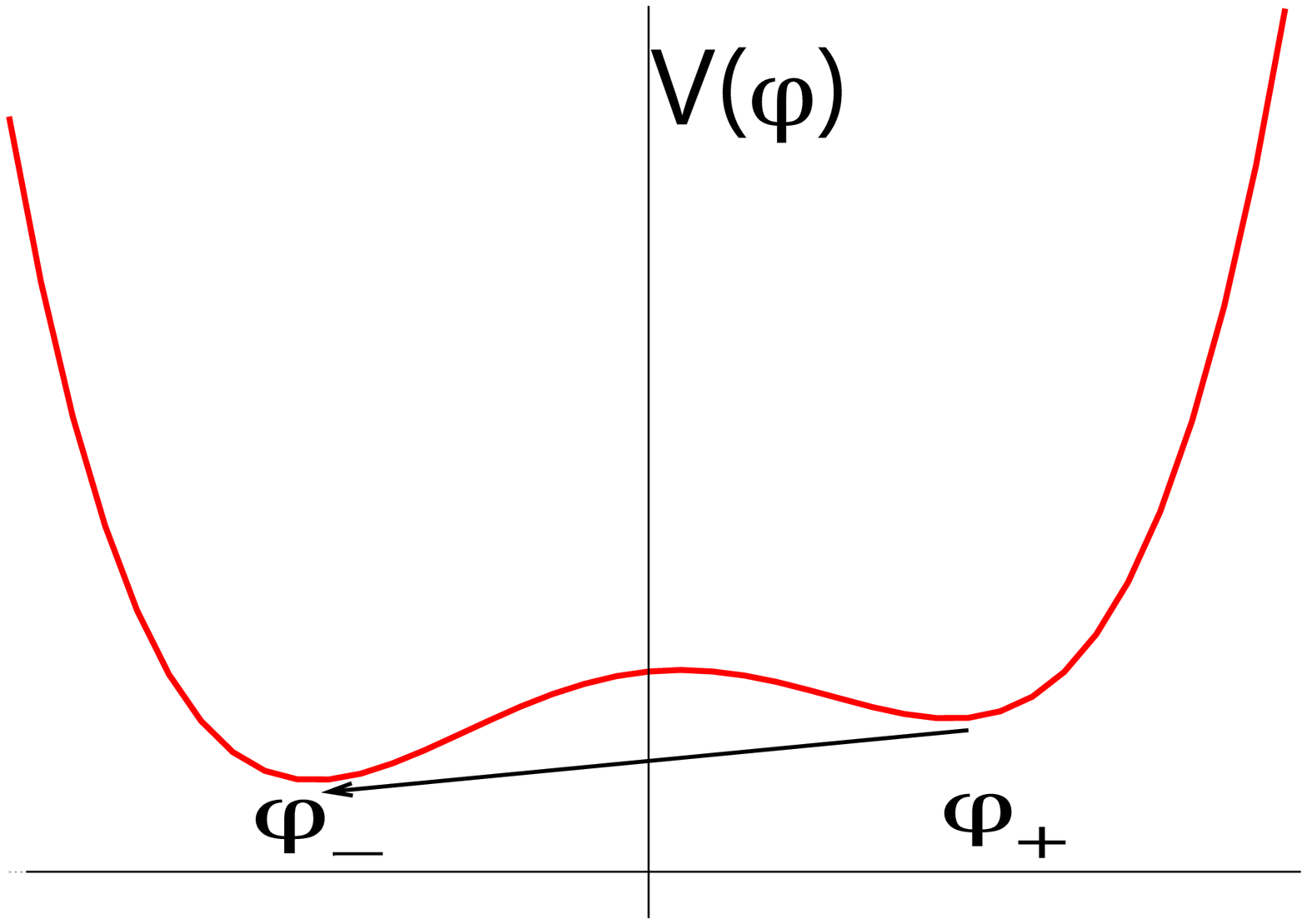} &
\hspace{0.5cm}
\includegraphics[width=5cm]{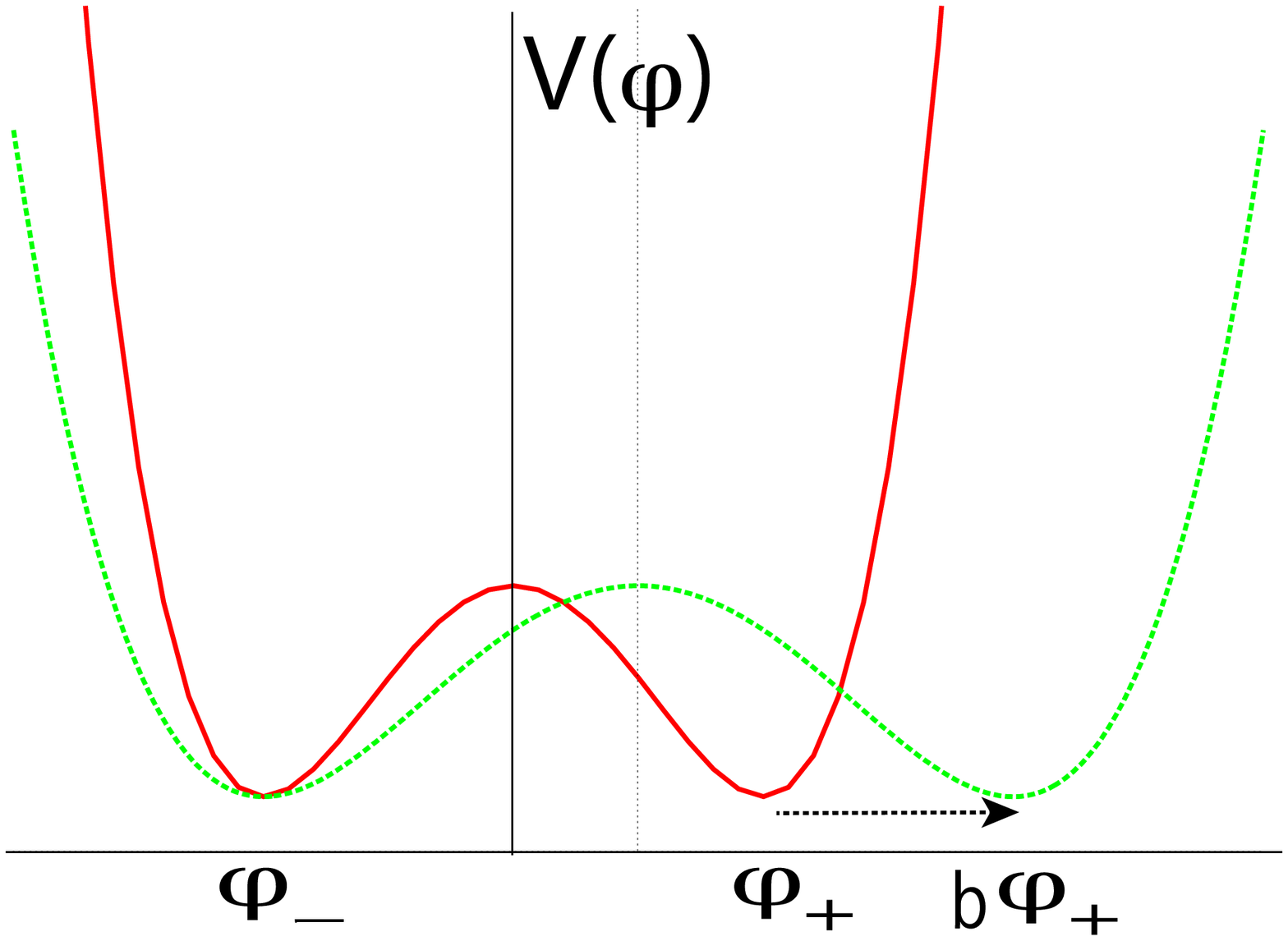} 
\end{tabular}
\end{center}
%\vspace{-1cm}
\caption{(Left) We plot the potential ${\cal V}(\Phi)$ in the Jordan frame.  
The tunnelling occurs from $\Phi_+$ to $\Phi_-$. (Center) We plot the potential 
$V(\phi)$ in the Einstein frame.  
The tunnelling occurs from $\phi_+$ to $\phi_-$. 
(Right) We plot the symmetric potential (\ref{eq:potential-wall}) 
and asymmetric one (\ref{eq:potential-wall2}). The positive vacuum  
$\phi_+$ where the local
minimum is located moves to the right direction by $b$-times  with the 
same height of the potential barrier.  }
\label{}
\end{figure*}
We will introduce the JBD theory \cite{Brans:1961sx,Fujii:2003pa} 
with a potential ${\cal V}(\Phi)$ 
in the Jordan frame 
as the model of varying of gravitational constant, which is expressed as the action  
\begin{equation}
S=\int d^4 x \sqrt{-g} \left[{\Phi^2\over 2} R-{1\over 2} \partial_\mu \Phi
 \partial^\mu \Phi -{\cal V}(\Phi)\right], 
\end{equation}
where $\Phi$ is the scalar field whose potential is considered 
to provide their semiclassical tunnelling, namely it has two minima called a 
true and false vacuum respectively. 
We use the units of $8\pi G_N=1$, where $G_N$ is the Newtonian gravitational constant. 

From such a potential, the bubble universe will be created as a standard instanton process. The gravitational constant $G$ 
is related to the vacuum expectation value (VEV) of the scalar field as follows
\begin{eqnarray}
G={G_N\over \langle\Phi\rangle^2}\,. 
\label{rel:G-Phi}
\end{eqnarray}
The scalar field has a different value depending on each potential minima, and 
then it can read different values of gravitational constants. 

In order to analyze this system, it is useful to study dynamics in 
the Einstein frame by using conformal 
transformation as $g_{\mu\nu}=\Phi^{-2}
\tilde{g}_{\mu\nu}$\cite{maeda1989}. After transformation, 
the action in the Einstein frame is given by 
\begin{equation}
S=\int d^4 x \sqrt{-\tilde{g}} \left[{\tilde{R}\over 2}-{1\over 2} \partial_\mu \phi
 \partial^\mu \phi -V(\phi)\right]\,. 
\end{equation}
Here we introduced a new scalar field: $\phi$ and its potential, which 
are related to 
the original ones in the Jordan frame as 
\begin{equation}
\phi=\phi_0+\sqrt{7}\ln\Phi,~~~V(\phi)={{\cal V}(\Phi)\over \Phi^4}
\,,
\label{rel:J-E-scalar-V}
\end{equation}
where $\phi_0$ corresponds to the VEV of the JBD scalar field 
in our universe ($\Phi_0=1$).
Hereafter we will omit the tilde which represents 
the variables in the Einstein frame  for brevity.  In this paper, as expected to 
generate a bubble, we consider the potential which should give tunnelling, 
which is expressed by
\begin{equation}
V(\phi)={\lambda\over 8}(\phi^2-\mu^2)^2+{\epsilon\over 2 \mu}
(\phi-\mu)+\Lambda\,.
\label{eq:potential-V}
\end{equation}
Under the approximation that $\epsilon$ is small, 
two vacua are obtained by 
$\phi_\pm=\pm\mu-{\epsilon\over  2\lambda \mu^3}+\calO(\epsilon^2)$ where 
the suffixes $\pm$ describe two minima. In the case where 
$\epsilon>0$, the tunnelling occurs 
from $\phi_+$ to $\phi_-$ vacuum. 
Each potential energy is given by 
$V(\phi_+)=
\Lambda+\calO(\epsilon^2)$ and $V(\phi_-)=\Lambda-\epsilon+\calO(\epsilon^2)$,
 respectively. 

 Since we assume that our universe was created by tunnelling from 
de Sitter universe, our vacuum state is given by $\phi_0=\phi_-$.
From the above potential (\ref{eq:potential-V}), 
the potential ${\cal V}(\Phi)$ in the original Jordan frame is found by 
 (\ref{rel:J-E-scalar-V}) .
%\begin{eqnarray}
%{\cal V}(\Phi)&=&\Phi^4
%\Bigl[{49 \lambda\over 8}\left((\ln\Phi)^2-{2\mu \over \sqrt{7}} 
%\ln \Phi
%\right)^2
%\nonumber \\
%&&
%~~~~+
%{\sqrt{7}\epsilon\over 2 \mu}\left(\ln\Phi-{2 \mu\over \sqrt{7}}\right)+\Lambda\Bigr]
%\,,
%\label{eq:potential-U}
%\end{eqnarray}
%where we set $\phi_0=\phi_-$ because our universe is in the true($-$) vacuum
%state. 
In Fig.1, we plot two potentials ${\cal V}(\Phi)$ and $V(\phi)$. From the figure, 
it is clear that the tunnelling process 
can be realized from a false($+$) vacuum to a true($-$) vacuum \footnote{Note that the possibility for tunnelling is found by use of 
the potential in the Einstein frame.
We may misunderstand the possibility if we look only at the potential in the Jordan frame.
See an example in the paper \cite{maeda1987}}.

%%%%%%%%%%%%%%%%%%%%%%%%%%%%%%%%%%%%%%%%%%%%%%%%%%%%%%%%%%%%%%%
\subsection{Nucleation of Bubble Universe}
We consider a spatially flat Freidmann-Lemaitre-Robertson-Walker (FLRW) spacetime
 in both sides of the wall. In this case, the instanton process is calculated in the Euclidean spacetime with $O(3)$ symmetry. 
 In the Lorentzian configuration, the line element takes the form 
\begin{equation}
ds^2=a^2(\eta)(-d\eta^2+dr^2+r^2 d\Omega^2)\,,
\end{equation}
where $\eta$ is a conformal time which is related to a cosmic time 
$t$ by $dt=a d\eta$. 

Following \cite{Basu:1991ig,KeskiVakkuri:1996gn,Simon:2009nb}, 
the calculation has been done by use of the 
complex time path approach, in which 
dynamics is studied by the complex time $\eta$. 
By adopting the thin-wall approximation, in which 
a thickness of wall is small enough 
compared to the size of a bubble, 
the bubble radius ${r}(\eta)$ is solved as a function of $\eta$.
The action is evaluated as
\begin{eqnarray}
S=&&\int d \eta\Bigl[{4\pi\over 3 } \epsilon a^4(\eta){r}^3(\eta)\notag\\
&&-4\pi \sigma a^3(\eta) {r}^2(\eta) \sqrt{1-(\partial_\eta{r}(\eta))^2)}
\Bigr]\,.
\end{eqnarray}
This action contains both contributions from  the volume and 
 surface area of three-dimensional sphere. 
Here we have used $\epsilon$ and $\sigma$, which
are the energy density difference between 
two vacua and the surface energy density of the wall, respectively. 
The equation of motion is given by taking a variation of the action $S$ 
with respect to ${r}$.  Solving it, we find a classical trajectory of the wall  ${r}(\eta)$. 
In the backward of time $\eta$, we trace a shrinking bubble radius and
 reach to a turning point where the canonical momentum 
${p}=\partial {\cal L}/\partial (\partial_\eta {r})$ vanishes. 
At this turning point $(\eta=\eta_i)$, a bubble nucleation   
occurs.
By the analytic continuation,
we find the wall trajectory in the complex $\eta$ plane, 
which shrinks  smoothly  to zero size of a bubble. 
 In order to perform this procedure, it is useful 
to rewrite the equation as one for $\eta({r})$. The task to do is reduced to 
solve it with the boundary conditions: 
\begin{equation}
{p}=4\pi \sigma 
{a^3 {r}^2 \over \sqrt{(\partial_{{r}}\eta)^2-1}}|_{\eta=\eta_i}=0\,,~~~
\partial_{{r}}\eta(0)=0\,.
\end{equation}
where $\eta=\eta_i$ denotes a nucleation time. 

The tunnelling rate per unit four-volume is obtained from the imaginary part of
 the action with 
the complex $\eta$, which becomes
\begin{equation}
\Gamma(\eta_i)\simeq \exp[-2\, {\rm Im} S(\eta_i)]\,.
\label{eq:Gamma-ImS}
\end{equation}
As a result \cite{Simon:2009nb}, the tunnelling probability   in a spatially  
flat de Sitter universe is obtained as 
\begin{equation}
{\rm Im}  S={\pi^2 \epsilon\over 3 H^4} \sinh^2\left[
{1\over 4} \ln(1+(3H \sigma/\epsilon)^2)\right]\,.
\label{eq:ImS}
\end{equation}
Note that the tunnelling rate is independent of the choice of $\eta_i$ in the de Sitter 
spacetime. This expression reduces to the result for Minkowski spacetime 
 known as the Coleman's bubble solution 
\cite{Coleman:1980aw} 
$${\rm Im}   S={27 \pi^2 \sigma^4\over  4\epsilon^3}\,,$$
in the limit of $H\to 0$.  
%%%%%%%%%%%%%%%%%%%%%%%%%%%%%%%%%%%%%%%%%
\subsection{Wall solution \label{wall solution}}
We  discuss how the difference of the VEV's $\phi_\pm$, 
 that is the difference of gravitational constants in the Jordan frame, affects
 the observational consequence. So  we analyze the dependence of the ratio 
$G_+/G_-=\Phi_-^2/\Phi_+^2=\exp[-{2\over \sqrt{7}}(\phi_+-\phi_-)]$ 
on the final result.
 For simplicity we discuss only the potential in the limit of $\epsilon\to 0$. 
That is the case of the potential for which both vacua have the 
same cosmological constant $\Lambda$
\footnote{We give an analytic solution of a domain wall for small $\epsilon$ in 
Appendix.
}.

The potential (\ref{eq:potential-V}) in this limit is
\begin{equation}
V_0(\phi)={\lambda\over 8}\left(\phi^2 -\mu^2\right)^2+ \Lambda
\,.
\label{eq:potential-wall}
\end{equation}
 This gives 
\begin{equation}
{G_+\over G_-}=\exp\left[-{4\mu\over \sqrt{7}}\right]
\,.
\end{equation}
So if we change the value of the parameter $\mu$, we can discuss the dependence of 
the different gravitational constants.
However, it also changes the height of the potential barrier. 
In order to keep the same potential height,
we adopt the following toy potential  
\begin{equation}
{V}_0(\phi)={2\lambda\over  (1+b)^4}(\phi- b \mu)^2(\phi+ \mu)^2+ \Lambda\,. 
\label{eq:potential-wall2}
\end{equation}
 The values of the scalar field in the mother universe and our universe are
given by $\phi_+=b\mu$ and $\phi_-=-\mu$, respectively. 
The ratio of the gravitational constants is 
\begin{equation}
{G_+\over G_-}=\exp\left[-{2(b+1)\mu\over \sqrt{7}}\right]\,.
\label{eq:G-b-phi}
\end{equation}

Comparing (\ref{eq:potential-wall}) with the case $b\neq 1$ in 
(\ref{eq:potential-wall2}) allows us to discuss the change of  
the ratio of $G_+/G_-$
with the same height of the potential barrier (see Fig.1). 
Calculating the wall solution and surface density of the wall 
$\sigma$  for this toy potential and varying the value of $b (>-1)$,
 we can see the effect of the different gravitational constant.

 In order to evaluate the surface energy density $\sigma$ of a bubble, we 
consider a spherical symmetric and static solution $\phi(r)$.
The basic equation is
\begin{equation}
\partial_r^2 \phi+{2 \over r}\partial_r \phi-{dV_0\over d\phi}=0
\,.
\end{equation}
Imposing the thin-wall approximation in which we can neglect 
the term of $\partial_r \phi$,   we can integrate it as 
\begin{equation}
\int_{\phi_i}^{\phi} {d\phi \over \sqrt{2 (V_0(\phi)-V_0(\phi_i))}}=r-r_i
\,,
\end{equation}
where $\phi_i=\phi(r_i)$.

For the potential  (\ref{eq:potential-wall2}), we can find the wall solution as 
\begin{eqnarray}
{\phi}(r)&=&-\mu+{(b+1)\mu\over 2}\left[1 +
\tanh \left({r-r_i\over 2d}\right) \right],
\label{sol:phi-r-wallb}
\end{eqnarray}
with
\begin{eqnarray}
d&=&{(1+b)\over2\mu\sqrt{\lambda} }
\,.
\end{eqnarray}
Of course,  in the limit of $b=1$, 
we find that  the solution (\ref{sol:phi-r-wallb}) 
is reduced to the
usual wall solution with the potential (\ref{eq:potential-wall}) 
\begin{equation}
\phi(r)= \mu \tanh\Bigl[{\mu \sqrt{\lambda}\over 2}(r-r_i)\Bigr]
\,.
\label{sol:phi-r-wall}
\end{equation}

Then we obtain $\sigma$ as
\begin{eqnarray}
\sigma= && \int_0^\infty 
dr \left[{1\over 2}(\partial_r \phi(r))^2+V(\phi)\right]\notag\\
\simeq && 
\int_{\phi_-}^{\phi_+} d\phi \sqrt{2(V_0(\phi)-V_0(\phi_-))}\,,
\end{eqnarray}
where we have used the thin-wall approximation. 
Substituting (\ref{eq:potential-wall2}) into the above equation,
we obtain the result 
\begin{equation}
\sigma(b)={\sqrt{\lambda}\mu^3\over 3 } (1+b)
\,.
\end{equation}
 When $b=1$, it gives the surface density of
 the original potential (\ref{eq:potential-wall})  as 
\begin{equation}
\sigma|_{b=1}={2\sqrt{\lambda}\over 3}\mu^3
\,.
\label{sol:sigma-1}
\end{equation}
 We can argue how
the difference of gravitational constants in the Jordan frame affects on 
the bubble nucleation rate $\Gamma$, the wall dynamics and 
the primordial perturbation through this surface density $\sigma(b)$,
which is a function of $b$. 

Note that we find the same surface density $\sigma$ even when 
we take into account of small $\epsilon$-modification of
the potential (see Appendix).

%%%%%%%%%%%%%%%%%%%%%%%%%%%%%%%%%%%%%%%%%%%%%%%%%
\section{Dynamics of bubble wall}
 In this section, we study 
dynamics of a bubble wall after nucleation of a bubble universe.
To obtain an analytic solution of a wall trajectory, we
assume that a wall is infinitely thin and 
impose a junction condition across the wall. 
The bubble dynamics has been studied by several authors 
\cite{Maeda:1985ye,Berezin:1987bc,Simon:2009nb,Casadio:2011jt,Ertan:2007eq}. 
  Following their works, we briefly summarize the results.

Each spacetime is distinguished by the suffixes $+$ and $-$, respectively. 
The bubble wall is represented as a timelike spherically symmetric hypersurface $\Sigma$ 
dividing a false ($+$) vacuum and  true ($-$) vacuum. 
 We adopt a spatially flat slicing of de Sitter universe for both spacetimes, 
whose metrics are given by 
\begin{equation}
ds_\pm^2=-dt_{\pm}^2+\exp(2 H_{\pm}t_{\pm} ) (dr_{\pm}^2+r^2_{\pm}d\Omega^2)\,,
\end{equation}
 where $H_{\pm}=\sqrt{\Lambda_{\pm}/3}$.  Note that 
\begin{equation}
H^2_+ -H^2_-=\epsilon/3\,.
\label{H+H-ep}
\end{equation}
When  $\epsilon\neq 0$, cosmological constants inside and outside of the wall
have different values. Up to the first order of $\epsilon$, they can be approximated as
$\Lambda_-=\Lambda-\epsilon$ and $\Lambda_+=\Lambda$, respectively. 
The metric of the bubble wall takes the form 
\begin{equation}
ds^2|_\Sigma=-d\tau^2+R^2d \Omega^2
\,.
\label{eq:metric-wall}
\end{equation}
We assume the stress-energy of this wall is given simply 
\begin{equation}
S_{\mu\nu}=-\sigma h_{\mu\nu}\,,
\end{equation}
 where  $h_{\mu\nu}=g_{\mu\nu}-n_\mu n_\nu$ is the projection tensor 
on the hypersurface $\Sigma$, which is 
the metric tensor of (\ref{eq:metric-wall}). 
The unit spacelike vector $n_{\mu}$ normal to the hypersurface $\Sigma$ 
is given by 
\begin{equation}
n_{\mu}=a\left(-{dr\over d\tau}, {dt\over d\tau}, 0, 0\right)\,,~
n^{\mu}=\left(a{dr\over d\tau}, {1\over a}{dt\over d\tau }, 0, 0\right)\,.
\label{eq:normal-vector-n}
\end{equation}
From 
$n^{\mu}n_{\mu}=1$, we obtain the relation equation 
\begin{equation}
\left({dt_{\pm}\over d\tau}\right)^2-a^2_{\pm}\left({dr_{\pm}\over d\tau}\right)^2=1\,.
\label{relation:t-r}
\end{equation}
By using ${dr\over d\tau}={dt\over d\tau} \times 
{dr\over dt}$, it can lead to 
\begin{equation}
{dt_\pm\over d\tau}={1\over \sqrt{1-a_\pm^2 ({dr\over dt_\pm})^2}}\,.
\label{eq:dt-dtau}
\end{equation}

%The Einstein equations can be written as the Gauss-Codazzi equations. 
%However, they are dealt with as constraint equations and we can solve the dynamics of wall% without using them. 
 In order to find the dynamical equation for the wall, we  
use the Israel's Junction conditions as 
\begin{eqnarray}
&&[g_{\mu\nu}]^+_-|_{\Sigma}=0\,,\label{cond;juntion1}\\
&&[K_{\mu\nu}]^+_-=-(S_{\mu\nu}-{1\over 2}h_{\mu\nu} S)\,,\label{cond;juntion2}
\end{eqnarray}
where $[X]^+_-=X|_+ -X|_-$ for the variable $X$ and $S=S^\mu_{~\mu}$. 
 The extrinsic curvature $K_{\mu\nu}$ of the hypersurface $\Sigma$ 
 is defined by 
\begin{equation}
K_{\mu\nu}=h^\rho_\mu h^\sigma_\nu \nabla_\sigma n_\rho\,,
\end{equation}
where 
 $\nabla_\sigma$ denotes a covariant derivative with respect to $g_{\mu\nu}$. 
 The continuity equation is given as 
\begin{equation}
S_\mu^{~\nu}{}_{||\nu}=-[S^{\rho\sigma}n_\rho h_{\mu\sigma}]^+_-
\,, 
\end{equation}
where $||$ denotes a covariant derivative with respect to 
the projection tensor $h_{\mu\nu}$. 
 This equation is reduced to the dynamical equation for 
 $\sigma$ as 
\begin{equation}
{d\sigma(\tau)\over d\tau}=\left[
a {dr\over d\tau} {d t\over d\tau}(\rho+P)\right]^+_-\,,
\label{eq:sigma-evol}
\end{equation}
where $\rho$ and $P$ denote the energy density and pressure in the bulk 
universes, respectively. 
If $\rho+P\neq 0$, e.g. for the radiation or matter dominant universe, 
$\sigma$ changes in time.
Throughout this paper, however, 
 since we consider 
only de Sitter background universes,
$\sigma$ turns out to be constant.  

From the junction condition (\ref{cond;juntion1}), we obtain 
the proper radius of the wall by
\begin{equation}
R=a_{+}r_{+}=a_{-}r_{-}\,.
\label{rel:R-ar}
\end{equation}
 To find the dynamical equation for $R$, 
we use the $(\theta\theta)$ component of the second junction 
equation (\ref{cond;juntion2}), which  leads to
\begin{equation}
K_{\theta\theta}^{+} -K_{\theta\theta}^{-}=-{\sigma\over 2}R^2\,,
\end{equation}
and its square becomes  
\begin{equation}
(K_{\theta\theta}^+)^2={1\over \sigma^2 R^4}\left\{(K_{\theta\theta}^-)^{2}
-(K_{\theta\theta}^+)^{2} -{\sigma^2 R^4\over 4}\right\}^2\,.
\label{Kthth2}
\end{equation}
While 
$K_{\theta\theta}$ is calculated as
\begin{equation}
K^2_{\theta\theta}=R^2\left\{1+\left({dR\over d\tau}\right)^2-H^2 R^2\right\}
\,.
\end{equation}
Using (\ref{Kthth2}), we can simplify the equation for $R$ as 
\begin{eqnarray}
\left\{{dR(\tau)\over d\tau}\right\}^2=B^2 R^2(\tau)-1\,,
\label{eq:diff-R}
\end{eqnarray}
where 
\begin{eqnarray}
B^2&=&H^2_{+}(1+c_{+}^2)=H^2_{-}(1+c_{-}^2)\,,
\label{defB}
\\
c_\pm&=&
H^{-1}_\pm \left({\epsilon\over 3\sigma}\mp{\sigma\over 4}\right)\,,
\label{eq:B^2}
\end{eqnarray}
 where the second equality in (\ref{defB}) is found from Eq. (\ref{H+H-ep}). 
For $c_+$, although $\epsilon$ is assumed to be small, 
 $c_+$ can be positive because 
\begin{equation}
{4\epsilon \over 3\sigma^2}=
{12 \over   (1+b)^2} \left({ \epsilon\over  \lambda \mu^4 }\right)
 \left({\mu\over m_{\rm PL}}\right)^{-2}\,,\label{eq:small-epsilon}
\end{equation}
can be smaller than unity,
if we assume $\mu\ll m_{\rm PL}$, 
where $m_{\rm PL}(=1)$ is the reduced Planck mass. 

For de Sitter bulk universes, which we have assumed here,
 $B$ is a constant and then a solution of a 
bubble radius $R(\tau)$ is obtained by 
\begin{equation}
R(\tau)={1\over B} \cosh B\tau\,.
\label{sol:wall-raduis}
\end{equation}

 We can also solve  the wall motion
in the interior or  exterior 
coordinates $(t_{+}, r_{+})$ or $(t_{-}, r_{-})$. 
In what follows, although 
we give the solution for the exterior coordinates,
the same form of the solution is obtained for 
the interior coordinates just  by replacement of $+$ with $-$.

From (\ref{rel:R-ar}), we have 
\begin{equation}
\ln R=H_+ t_+ +\ln r_+\,,
\label{rel:lnR-t-r}
\end{equation} 
and differentiate both sides of the equation with respect to $\tau$. 
Taking its square and  using 
(\ref{relation:t-r}) and (\ref{sol:wall-raduis}), we find the equation for
 ${d(\ln r_+)/d\tau}$ as 
\begin{equation}
{d(\ln r_+)\over d\tau}={{\partial_\tau(\ln R)}\pm H^2_+ R |c_+| \over 1-H^2_+ R^2}
\,,
\end{equation}
which is integrated by using the solution of (\ref{sol:wall-raduis}) as
\begin{equation}
r_+={r_{\infty+} \cosh B\tau\over |\sinh B\tau \pm c_+|}\,,
\label{sol:r-tau}
\end{equation}
where $r_{\infty+}$ is an integration constant to be determined later and 
$\pm$ take $+$($-$), 
if the sign of $c_+$  takes 
a positive (negative) value. 
Inserting (\ref{sol:r-tau}) into (\ref{rel:lnR-t-r})  
 leads to the solution of $t_{+}$ as 
\begin{equation}
H_{+} t_+=\ln (|\sinh B\tau \pm c_+ |)-\ln B r_{\infty +}\,.
\label{sol:t-tau}
\end{equation}

Finally we  solve the wall motion  as  $r_+(t_+)$, which
 denotes a dynamics of the comoving bubble radius $r$ for the observer in the 
exterior universe. 

Combining ${dR\over d\tau}={dt\over d\tau}\times {d\over dt }(a{r})$, (\ref{eq:dt-dtau}) 
and (\ref{sol:wall-raduis}), 
we have a differential equation
\begin{equation}
{-B^2 R^2 \over H_+}{\partial_t {r}(t_+)\over {r}}=
1 -|c_+|\sqrt{B^2 R^2-1}\,.
\end{equation}
It is integrated when rewritten as equation for $B^2 R^2$ as 
\begin{equation}
{r}_+(t_+)=\sqrt{(a^{-1}_+-1)^2+c^{-2}_+}H^{-1}_+\,.
\label{r+}
\end{equation}
Here we take a nucleation time of a bubble as $t_+=0$, when 
we impose $\partial_t {r}_+(0)=0$. 
The solution takes the same form for both interior coordinates too. 

We can summarize dynamics of bubble wall in the 
inside and outside observer as follows. 
At the initial time $t_\pm=0$, the bubble 
is created at the comoving radii
\begin{equation}
{r}_\pm(0)={r}_{i\pm}=\left
|{\epsilon\over 3\sigma}\mp {\sigma\over 4}\right|^{-1},
\end{equation}
 and then expands.
The comoving radii eventually  converge to 
\begin{equation}
{r}_\pm(t\to \infty)=\sqrt{1+c^2_\pm}{r}_{i\pm}\,
\equiv r_{\infty\pm}\,.
\end{equation}
The integration constant $r_{\infty+}$ given in (\ref{sol:r-tau})  
should be the same as the above value.
The asymptotic comoving radii are rewritten by 
\begin{equation}
r_{\infty\pm}=\sqrt{H^{-2}_\pm+(r_{i\pm})^2}\,.
\label{eq:r0pm}
\end{equation}
If we ignore small initial radii $r_{i\pm}$, this equation basically 
yields $r_{\infty\pm}\sim H^{-1}_\pm$, which means the final convergent radii are
approximately described by the Hubble radii $H^{-1}_\pm$ in each coordinates. 
It is obviously noted  that physical scale of the bubble radius continues to expand 
as $R=a_\pm r_\pm
\propto e^{H_\pm t_\pm}$, 
although the comoving radii converge to about the Hubble radii. 
%%%%%%%%%%%%%%%%%%%%%%%%%%%%%%%%%%%%%%%%%%%%%%%%%%%%%%%
\section{Perturbations \label{perturbations}}
In this section, in order to discuss observational effects of the different gravitational 
constant in the mother universe, we analyze metric perturbations in 
the Newtonian (longitudinal) gauge. The perturbed metric  is given by
\begin{equation}
ds^2=a^2[-(1+2\psi)d\eta^2+(1-2\psi)(dr^2+r^2 d\Omega^2)]\,.
\end{equation}
In the interior and exterior 
spacetimes, the tunnelling JBD scalar field takes the values of
$\phi_-$ and $\phi_+$, respectively, which provide 
non-zero cosmological constants. As a result, each bulk universes expand
exponentially as  de Sitter spacetime. 
However, such a de Sitter expansion will not end.
 In order to discuss more realistic cosmological scenario, 
instead of a cosmological constant,
we  introduce another scalar field $\varphi$, which is confined 
at our  vacuum state $\phi=\phi_-$. This scalar field  $\varphi$ is responsible 
for inflation and will finish the exponential expansion. 
The quantum fluctuation of this scalar field provides the density perturbations,
which we will discuss here \footnote{In the outside mother universe, we may also find an
inflaton, which will finish the exponential expansion.
In our analysis of perturbations, however, we assume de Sitter expansion 
in both bulk universes.  Hence our result does not change 
unless the outside inflation will end earlier than our inflation. 
There may be some effects on the perturbations from the inflaton field,
which we ignore in our analysis.}.

The new potential 
$U(\varphi)$ is assumed to be  independent of $V(\phi)$. In this case, we find 
the background equations as 
\begin{eqnarray}
&&{\cal H}^2= {1\over 3}\left({1\over 2}\varphi'^2 + a^2 U\right)\,,\\
&&{\cal H}'-{\cal H}^2=-{1\over 2}\varphi'^2\,,\\
&&\varphi''+2{\cal H}\varphi'+ a^2 U_{,\varphi}=0
\,,
\end{eqnarray}
 where 
${\cal H}=Ha$ and a prime denotes a derivative with respect to 
a conformal time $\eta$.
We assume a slow-roll inflation for $\varphi$ field for simplicity, 
i.e. 
\begin{align}
a(t)=\exp(Ht),~~~{\cal H}=-{1\over \eta},~~-{\dot{\varphi}^2\over 2}=
\dot{H}
\,,
\end{align}
 where a dot  denotes a derivative with respect to 
the cosmic time $t$. 
We  introduce 
slow-roll parameters as 
\begin{equation}
\varepsilon_1=-{\dot{H}\over H^2},
~~\varepsilon_2={\dot{\varepsilon_1}\over H \varepsilon_1}
\,.
\end{equation}
Adding the perturbation of scalar field $\delta \varphi$ and expanding the basic equations 
up to linear order, we find one master equation by use of the 
Mukhanov-Sasaki variable $v$, which is a linear combination of $\delta \varphi$ 
and $\psi$.
Defining 
\begin{equation}
u={a\psi\over \varphi'},~~~v=a\left(\delta\varphi+{\varphi'\over H}\psi\right)
\,,
\end{equation}
we find  the basic equations for these variables  as 
\begin{equation}
\Delta u=z(v/z)'\,,~~v=\theta(u/\theta)'
\,,
\end{equation}
where we have used
\begin{equation}
\theta={{\cal H}\over a\varphi'}={1\over a \sqrt{2\varepsilon_1}}\,, ~~z={1\over \theta}\,.
\end{equation}
Then we obtain the closed master equation for $u$: 
\begin{equation}
u''-\left(\Delta +{\theta''\over \theta}\right)u=0
\,.
\label{eq:differ-u}
\end{equation}

Under a slow-roll condition, the term $\theta''/\theta$ is evaluated as
\begin{equation}
{\theta''\over \theta}={1\over \eta^2}\left(\nu^2-{1\over 4}\right)\,,~~~
\nu^2={1\over 4}+\varepsilon_1+{\varepsilon_2\over 2}\,.
\end{equation}

For spherically symmetric perturbations, a solution $u(r,\eta)=e^{\pm i k r}/r$ satisfies the equation 
$$\Delta u=(\partial_r^2+2\partial_r /r)u=-k^2 u\,.$$
The mode function with a comoving wave number $k$ is obtained by 
\begin{equation}
u={e^{\pm i kr}\over r}\sqrt{-\eta}\left[C_1 H_\nu^{(1)} (-k \eta) +C_2 H_\nu^{(2)} 
(-k \eta)\right]\,,
\end{equation}
where $C_{1,2}$ are arbitrary constants. 
In the interior coordinate($-$) system,  $v$ 
is reduced to $v_k\to e^{-i k\eta}/\sqrt{2k}$ in the high-frequency limit,
which corresponds to an adiabatic vacuum. 
This condition fixes the above constants  as $C_{1,2}$.  
Then the solution $u$ is determined up to a phase factor as
\begin{equation}
u_{\rm in}={e^{i k_-r_-}\over r_-}{\sqrt{-\pi \eta_-}\over 2k_-}
 H_\nu^{(1)} (-k_- \eta_-)\,.
\label{sol:u-in}
\end{equation}

This mode solution is proportional to 
 $e^{ik_-(r_--\eta_-)}$ in the large $|\eta_-|$ limit, so interpreted as  an incoming wave
from the past. 
It propagates in the outward (large $r_-$) direction from inside the bubble 
and then approaches to the boundary wall. 

This incoming wave is scattered at the wall  and then divided into reflected and 
transmitted waves, which are written by 
\begin{eqnarray}
&&u_{\rm rf}={e^{-i k_-r_-}\over r_-}{\sqrt{-\pi \eta_-}\over 2k_-}
 H_\nu^{(1)} (-k_- \eta_-)\,,\\
&&u_{\rm tr}={e^{i k_+ r_+}\over r_+}
{\sqrt{-\pi \eta_+}\over 2k_+}
 H_\nu^{(1)} (-k_+ \eta_+)
\,,
\end{eqnarray}
respectively. 
Here the outside wavenumber: $k_+$ is related to $k_-$ as 
$k_+=(a_+/a_-) k_-$ because the phase factor $e^{i k(r\pm \eta)}$ and 
the proper time $d\tau^2=a^2_\pm(d\eta_\pm^2-dr^2_\pm)$ have to be 
invariant across the wall. 

The  scattered mode $u_{\rm rf}$ is moving 
in the inward direction and then  
is going back to the interior bubble. 
While the  mode $u_{\rm tr}$ is transmitted  
to the outside bubble (the exterior universe). 

Expressing each perturbation as 
\begin{equation}
u_-=u_{\rm in}+\beta u_{\rm rf},~~~u_+=\alpha u_{\rm tr}
\,,
\end{equation}
we match those wave solutions at the hypersurface by use of
 the junction condition.

The matching has to be done  for 
each proper time of the wall $\tau$.
We then impose that the wave amplitudes and their
 normal derivatives are continuous at the wall hypersurface as
\begin{align}
&u_-(\tau)=u_+(\tau)\,,\label{eq:matching-cond0}\\
&n^{\mu}_- \partial_\mu u_-(\tau)=n^\mu_+ \partial_\mu u_+(\tau)\,.
\label{eq:matching-cond}
\end{align}
The similar argument 
was given in \cite{Ertan:2007eq} as the 
matching perturbation across a hypersurface. 

\begin{figure*}[t]
\vspace{-1cm}
\begin{center}
\begin{tabular}{ll}
\includegraphics[width=6cm]{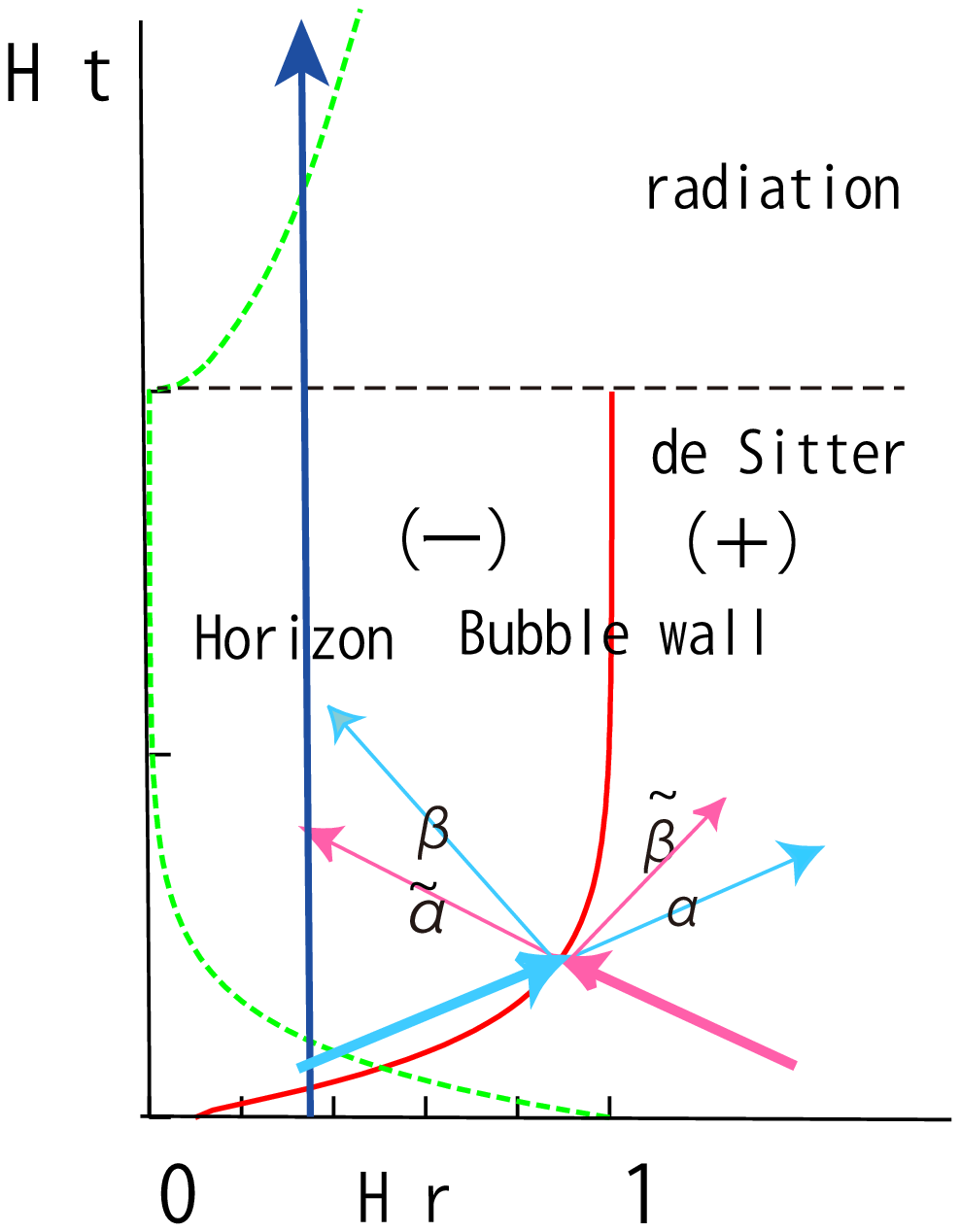} & 
\hspace{2cm}
\includegraphics[width=7cm]{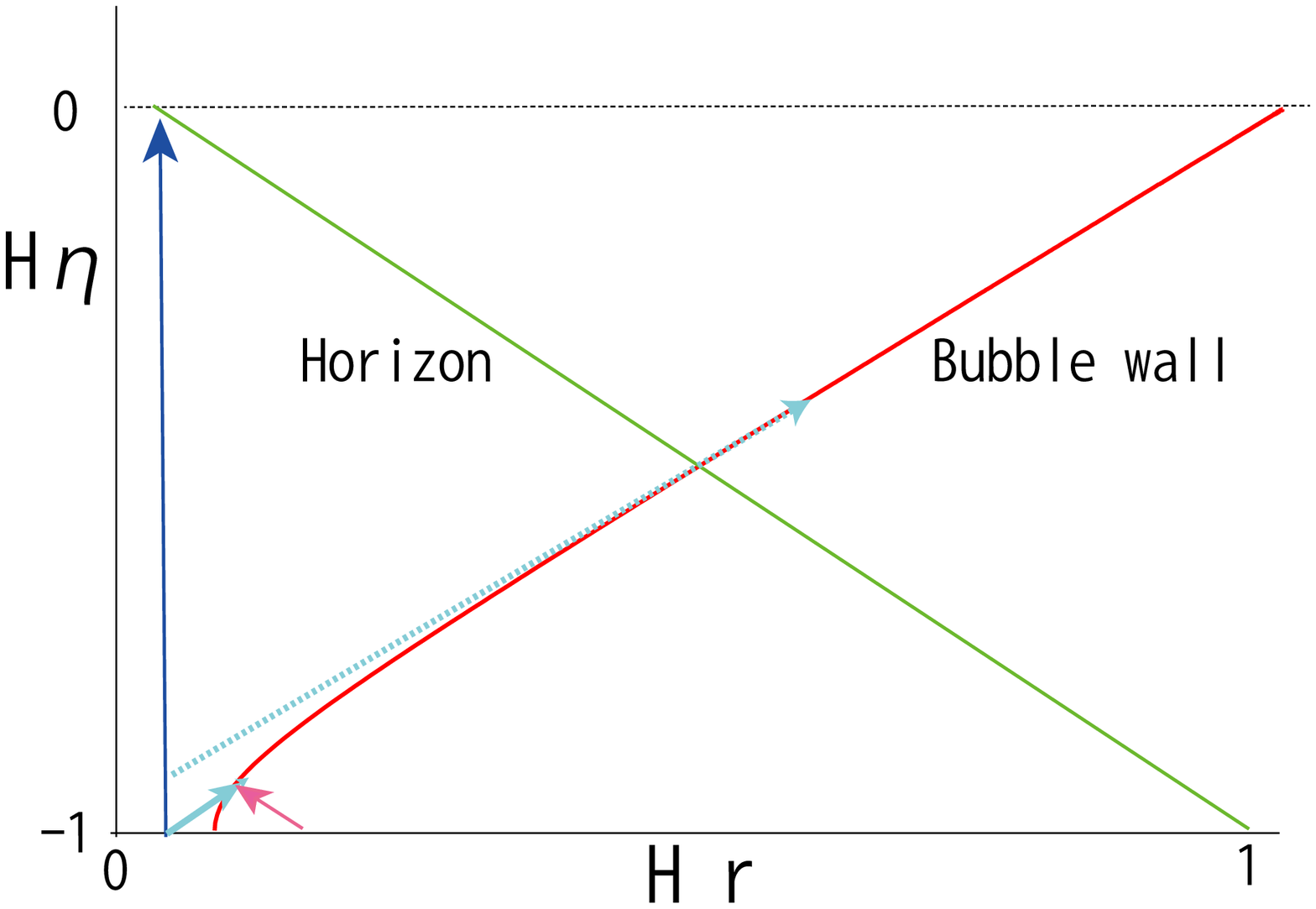}
\end{tabular}
\end{center}
%\vspace{-1cm}
\caption{We plot the trajectories representing the bubble wall (red) and 
horizon scale (green) 
as seen by a observer outside the wall in the units of 
$H_-=1$.  The blue line denotes a typical perturbation scale, which is in superhorizon scale during de 
Sitter phase. It is shown by the light-blue lines 
that an incoming wave propagates from inside the bubble to the boundary wall, 
scattered at the wall and then divided into a reflected $\beta$ and transmitted $\alpha$ waves. The pink line denotes incoming wave from the exterior bubble and scattered into a reflected $\tilde{\beta}$ and transmitted 
$\tilde{\alpha}$ waves similar as light-blue lines.  
(Left) We plot them in the coordinate $t$ and comoving radius $r$.  
(Right) We plot them in the coordinate conformal time $\eta=-1/(aH)$ 
and comoving radius $r$. It is shown that the slope of incoming wave (light-blue) is an angle of $45^\circ$. So at late time, the reflection is not 
generated since the slope of incoming asymptotically becomes a same 
slope of the wall trajectory, i.e., $\beta\simeq 0$ and similarly 
all waves from the exterior bubble  
can be transmitted, i.e., $\tilde{\alpha}\simeq 1$.  }
\label{Fig-TR}
\end{figure*}
 From the wall motion 
(\ref{sol:r-tau}) and (\ref{sol:t-tau}) with (\ref{r+}), 
the normal vector $n^{\mu}$  (\ref{eq:normal-vector-n}) is found as 
\begin{equation}
n^t_\pm={|c_\pm|\sinh B\tau-1\over \sinh B\tau +| c_\pm|}\,,~~
n^r_\pm={B^2 r_{\infty\pm} \cosh B\tau \over H_\pm  (\sinh B\tau +|c_\pm|)^2}\,.
\end{equation}
Rewriting $a_\pm$ and $\eta_\pm$ as a function of $\tau$ and 
using the matching conditions (\ref{eq:matching-cond0}) and (\ref{eq:matching-cond}), 
we obtain the amplitudes of the reflected and transmitted waves. 
We plot some wave trajectory in Fig.\ref{Fig-TR}.

In the comoving radial coordinate $r$, the bubble radius (see the red line in Fig.\ref{Fig-TR}) 
increases from an initial value $r_i$ 
to a convergent value $r_{\infty\pm}\simeq 1/H_\pm$, while 
a comoving horizon $1/(H_\pm a_\pm)$ (the green line) 
decreases monotonically from an initial value 
$1/H_\pm$ where we have set $a_i=1$ at an initial time 
$t=0$   and approaches asymptotically to zero 
during de Sitter expansion.  However 
if de Sitter expansion continues forever, 
 a typical perturbation mode (as written in a blue line in Fig.\ref{Fig-TR}) keeps to be 
located in superhorizon scale and cannot enter in the horizon again, 
which is not observable.  
Such a model does not 
describe a realistic scenario for cosmological perturbations. 

Hence as discussed above, we have added  another scalar field $\varphi$, 
which is responsible for inflation.
In a realistic inflationary scenario, 
inflation will end and lead to reheating of the Universe,
 connecting to a usual radiation dominated FLRW universe. Taking 
such a realistic scenario into account, we find 
the horizon scale increases 
again from the convergent value, and then such superhorizon 
perturbation will reenter into the horizon. 
The initial incoming wave of 
perturbations 
 approaches to the wall, which is located on over the horizon 
(see the solid light-blue line with an arrow), 
and then is scattered into the reflected  and transmitted waves. 
Now we  shall analyze how 
this reflection mode affects the observed quantities, i.e. 
the CMB power spectrum  of the perturbations by the observer lived in the interior 
bubble universe.

Generally speaking, if one of 
the bulk spacetimes  is not de Sitter, 
it is difficult to find an analytic solution
for the perturbations as well as the background 
spacetime.
We need a complicated numerical analysis. 
However, since the universe expands exponentially during 
inflation, 
we assume the simplest situation for the calculation of 
the wave propagation, that is, the case 
of a de Sitter expansion in both sides of spacetime. 

In a de Sitter background, we can calculate 
the reflection amplitude $\beta$ and discuss 
how this mode affects on perturbations. 
We match perturbations in both sides at each proper 
time $\tau$. Eliminating $\alpha$ in the junction conditions 
(\ref{eq:matching-cond0}) 
and (\ref{eq:matching-cond}), we 
find $\beta$ as 
\begin{widetext}
\begin{align}
\beta=&e^{2 i k_- r_-}\times\Bigl[-\Bigl(f_+(\tau)-f_-(\tau)\Bigr)H_{\nu}^{(1)}(-k_+\eta_+)
H_\nu^{(1)}(-k_-\eta_-)\notag\\
&
+g_+(\tau)H_{\nu+1}^{(1)}(-k_+\eta_+)
H_\nu^{(1)}(-k_-\eta_-)
%\notag\\
%&
- g_-(\tau)H_{\nu+1}^{(1)}(-k_-\eta_-)
H_\nu^{(1)}(-k_+\eta_+)\Bigr]\nonumber\\
&\times \Bigl[\Bigl(f_+(\tau)-f^*_-(\tau)\Bigr)H_{\nu}^{(1)}(-k_+\eta_+)
H_\nu^{(1)}(-k_-\eta_-)\notag\\
&
-g_+(\tau)H_{\nu+1}^{(1)}(-k_+\eta_+)
H_\nu^{(1)}(-k_-\eta_-)
%\notag\\
%&
+ g_-(\tau)H_{\nu+1}^{(1)}(-k_-\eta_-)
H_\nu^{(1)}(-k_+\eta_+)
\Bigr]^{-1}\,, 
\label{sol:beta}
\end{align}
where $*$ denotes a complex conjugate and 
\begin{eqnarray}
&& f_\pm(\tau)=\Bigl[(|c_\pm|\sinh B\tau -1){H_\pm(1+2 \nu) \over 2 B}
%\nonumber\\
%&&
 + (1-ik_\pm r_{\infty\pm} y_\pm(\tau)^{-1}\cosh B\tau ) {B\over H_\pm} \Bigr] y_\pm(\tau)^{-1}\, ,\label{eq:f-g-eta-y0}\\
&& g_\pm(\tau)=k_\pm r_{\infty\pm}(|c_\pm| \sinh B\tau -1) y^{-2}_\pm(\tau)\,,
\\
&& -k_\pm \eta_\pm(\tau)={k_\pm Br_{\infty\pm} \over H_\pm y_\pm(\tau)}\,,
\label{eq:f-g-eta-y}
\end{eqnarray}
with
\begin{eqnarray}
y_\pm(\tau)=\sinh B\tau + |c_\pm|\,.
\end{eqnarray}
We note that in the above calculation we have used 
the same value of $\nu=1/2$ for the Hankel functions in both sides of the wall. 
Strictly speaking the order of the Hankel function has to be  $\nu_\pm$, 
which depend on the slow-roll parameters $\varepsilon_1, \varepsilon_2$.
 In our calculation,  however, both sides of spacetime are approximated by de Sitter, 
that is, slow-roll parameters vanish and then we 
approximate the wave functions with $\nu_\pm=\nu=1/2$. We 
have also used the formula of the Hankel functions: 
 $\partial_z H_\nu(z)=\nu z^{-1}H_\nu(z)-H_{\nu+1}(z)$.

Using the definition of $H_{1/2}(z)$ and $H_{3/2}(z)$, we find 
\begin{align}
\beta=e^{2 i k_- r_-}\times
{f_+(\tau)-f_-(\tau)+{g_+(\tau)\over k_+\eta_+}-{g_-(\tau)\over k_-\eta_-}
+i[g_+(\tau)-g_-(\tau)]
\over 
-(f_+(\tau)-f^*_-(\tau))-{g_+(\tau)\over k_+\eta_+}+{g_-(\tau)\over k_-\eta_-}
-i[g_+(\tau)-g_-(\tau)]}\,.
\label{sol:beta2}
\end{align}

If $\epsilon$ vanishes exactly, i.e., 
there is no energy difference between two vacuum states, 
since  $H_+=H_-$, 
$f_+=f_-$, $g_+=g_-$, and  $\eta_+=\eta_-$, 
 then we find $\beta=0$. 
However, we assume $\epsilon/3=H_+^2-H_-^2\neq 0$, 
which results in  
$\beta\propto \epsilon \neq 0$ if 
$\epsilon$ is small. 
In fact, assuming $\epsilon\ll \lambda \mu^4\times (\mu/m_{\rm PL})^2$ as well as
$\epsilon\ll \Lambda$, we find 
\begin{eqnarray}
\beta&=&e^{2 i k_- r_0}\times{c_0\over \sqrt{1+c_0^2} 
\cosh(B_0\tau)}{\epsilon\ell^2\over 12}
\Big\{{i\over k_-\ell }\left(\sinh(B_0\tau)-c_0^{-1}\right)
\notag\\
&&
+{\sqrt{1+c_0^2}\over c_0}\cosh(B_0\tau)\left(1 +{\eta_0\over \ell}\right)
-\sinh(B_0\tau)\left(1+{1\over c_0^2} +{\eta_0\over \ell}
\right)+{\eta_0\over c_0\ell }
\Big\}
\,,
\end{eqnarray}
where we define 
\begin{eqnarray}
\ell&=&\sqrt{3\over \Lambda}=H_+^{-1}
\,,~~
c_0={\ell\sigma\over 4}
\,,~~r_0={\ell \sqrt{1+c_0^2}\over c_0}\,,\nonumber\\
B_0&=&{\sqrt{1+c_0^2}\over \ell}
\,,~~
\eta_0=-{(1+c_0^2)\ell \over c_0(\sinh(B_0\tau)+c_0)}
\,.
\end{eqnarray}
\end{widetext}

Similarly, we also have to estimate the contribution from the perturbation
 in the exterior universe (see the pink line in Fig.\ref{Fig-TR}). Now the incoming wave  $u$ is given as
\begin{equation}
u_{\rm in}={e^{-i k_+r_+}\over r_+}{\sqrt{-\pi \eta_+}\over 2k_+}
 H_\nu^{(1)} (-k_+ \eta_+)\,.
\label{sol:u-in}
\end{equation}
It is scattered at the wall  and then divided into reflected and 
transmitted waves, which are written by 
\begin{eqnarray}
&&u_{\rm rf}={e^{i k_+r_+}\over r_+}{\sqrt{-\pi \eta_+}\over 2k_+}
 H_\nu^{(1)} (-k_+ \eta_+)\,,\\
&&u_{\rm tr}={e^{-i k_- r_-}\over r_-}
{\sqrt{-\pi \eta_-}\over 2k_-}
 H_\nu^{(1)} (-k_- \eta_-)
\,,
\end{eqnarray}
respectively. 
Expressing each perturbation as 
\begin{equation}
u_+=u_{\rm in}+\tilde{\beta} u_{\rm rf}\,,~~~u_-=\tilde{\alpha} u_{\rm tr}
\,,
\end{equation}
we match those wave solutions at the hypersurface by use of
 the junction condition. 

Eliminating $\tilde{\beta}$ in the junction conditions 
(\ref{eq:matching-cond0}) 
and (\ref{eq:matching-cond}), we 
find the transmitted waves of the perturbations in the exterior universe, 
that is moving inward direction to the interior universe. The transmitted rate $\tilde{\alpha}$ is given by 
\begin{widetext}
\begin{align}
&\tilde{\alpha}={\sqrt{\eta_+ \over \eta_-}}{k_- r_-\over k_+r_+}
e^{i (k_- r_- -k_+ r_+)}\times
\Bigl[\Bigl(f_+(\tau)-f_+^*(\tau)\Bigr)(H_{\nu}^{(1)}(-k_+\eta_+))^2\Bigr]
\notag\\
&\times \Bigl[\Bigl(f_+(\tau)-f^*_-(\tau)\Bigr)H_{\nu}^{(1)}(-k_+\eta_+)
H_\nu^{(1)}(-k_-\eta_-)
-g_+(\tau)H_{\nu+1}^{(1)}(-k_+\eta_+)
H_\nu^{(1)}(-k_-\eta_-)
\notag\\
&
+ g_-(\tau)H_{\nu+1}^{(1)}(-k_-\eta_-)
H_\nu^{(1)}(-k_+\eta_+)
\Bigr]^{-1}.
\label{}
\end{align}

Using the definition of $H_{1/2}(z)$ and $H_{3/2}(z)$, we find 
\begin{align}
\tilde{\alpha}=\sqrt{k_-\over k_+}{H_-\over H_+}
e^{i (k_- (r_-+\eta_-) - k_+(r_+ + \eta_+))}\times
{f_+(\tau)-f_+^*(\tau)\over 
(f_+(\tau)-f^*_-(\tau))+{g_+(\tau)\over k_+\eta_+}-{g_-(\tau)\over k_-\eta_-}
+i[g_+(\tau)-g_-(\tau)]}\,.
\label{}
\end{align}

When
$\epsilon$ is small, we find 
\begin{eqnarray}
&&\tilde{\alpha}=1+{\epsilon\ell^2\over 12}
\Biggl[2 i k_- \eta_0+
{c_0\over \sqrt{1+c_0^2} 
\cosh(B_0\tau)}
\Bigg\{{i\over k_-\ell }\left(\sinh(B_0\tau)-c_0^{-1}\right)
\notag\\
&&-{\sqrt{1+c_0^2}\over c_0}\cosh(B_0\tau)\left(4-{1\over c_0^2} +{2\eta_0
\over \ell}\right)
-\sinh(B_0\tau)\left(1+{1\over c_0^2} +{\eta_0
\over \ell}
\right)+{\eta_0
\over c_0\ell}
\Bigg\}\Biggr]
\,.\label{eq:til-alpha-e}
\end{eqnarray}
\end{widetext}

Once we would obtain the solution of $u$, we have to convert it to 
the curvature perturbation $\zeta$ and estimate its power spectrum. 
In Fourier space,  we find $\zeta_k=v_k/z=\theta^2(u_k/\theta)'$. 

Hence the curvature perturbation is given by
\begin{widetext}
\begin{eqnarray}
\zeta_k=&&
 {\sqrt{-\pi \eta_-}\over 2k_-}e^{- i k_- r_-} 
\Biggl\{H_{1/2}^{(1)}(-k_-\eta_-)
\Bigl[{H_-\over \sqrt{2\varepsilon_{1-}}}
\left({\varepsilon_{2-}\over 2}\beta-{{d\beta\over d\tau}\over {d\eta_-\over d\tau}}
\eta_-\right)+
 {H_+\over \sqrt{2\varepsilon_{1+}}}
\left(\left(1+{\varepsilon_{2+}\over 2}-{\eta_+\over \eta_-}{{d\eta_-\over d\tau}\over {d\eta_+\over d\tau}}\right)\tilde{\alpha}-{{d\tilde{\alpha}\over d\tau}\over {d\eta_+\over d\tau}}\eta_+\right) \Bigr]\notag\\
&&-k_- \eta_- H_{3/2}^{(1)}(-k_-\eta_-)
\Bigl[{H_-\over \sqrt{2\varepsilon_{1-}}}\beta
+ {H_+\over \sqrt{2\varepsilon_{1+}}}
{\eta_+\over \eta_-}{{d\eta_-\over d\tau}\over {d\eta_+\over d\tau}}
\tilde{\alpha}\Bigr]\Biggr\}\,.
\end{eqnarray}
Using the definition of Hankel function, 
a dimensionless power spectrum can be 
obtained by 
\begin{eqnarray}
&&\Delta_\zeta^2(k_-)\equiv {k_-^3\over 2\pi^2}|\zeta_k|^2\notag\\
&&=
{1\over 8 \pi^2 }
\Bigg|(1+i k_- \eta_-){H_-\over \sqrt{\varepsilon_{1-}}}
\beta+\left(1+i k_-\eta_- {\eta_+\over \eta_-}{{d\eta_-\over d\tau}
\over {d\eta_+\over d\tau}}\right)
{H_+ \over \sqrt{\varepsilon_{1+}}}\tilde{\alpha}
-{{\eta_-}\over {d\eta_-\over d\tau }}{H_-\over \sqrt{\varepsilon_{1-}}}
{d\beta\over d\tau}-{\eta_+\over {d\eta_+\over d\tau}}
{H_+ \over \sqrt{\varepsilon_{1+}}}
{d\tilde{\alpha}\over d\tau}\Bigg|^2\,,
\label{eq:modulation}
\end{eqnarray}
where we neglected a slow-roll correction, i.e., $\varepsilon_{2\pm}=0$. 

We can divide the deviation from the perturbations in a standard inflationary model 
into three parts $s_0, s_1$ and $s_2$, which are of order $\calO(\epsilon)$, as 
\begin{eqnarray}
8\pi^2 \ell^2 \varepsilon_{1+}\Delta_\zeta^2(k_-)
\equiv1+s_0+s_1+s_2+\calO(\epsilon^2)
\,,
\label{eq:def-dimpower}
\end{eqnarray}
with 
\begin{eqnarray}
s_0 &=&
2  
{\rm Re}\left( \sqrt{\varepsilon_{1+}\over \varepsilon_{1-}}
\beta+ (\tilde{\alpha}-1)\right)
\notag\,,\\
s_1&=
&
2  
{\rm Re}\left(-{{\eta_-}\over {d\eta_-\over d\tau }}
\sqrt{\varepsilon_{1+}\over \varepsilon_{1-}}
{d\beta\over d\tau}-{\eta_+\over {d\eta_+\over d\tau}}
{d\tilde{\alpha}\over d\tau}\right)
\, 
,\notag\\
s_2&=&
2 
i k_- \eta_-\times {\rm Im}\left(\sqrt{\varepsilon_{1+}\over \varepsilon_{1-}}
\beta+  {\eta_+\over \eta_-}{{d\eta_-\over d\tau}
\over {d\eta_+\over d\tau}}
 (\tilde{\alpha}-1)\right)
\,.
\label{eq:def-dimpower1}
\end{eqnarray}
where Re and Im denote a real and imaginary part, respectively. 

At a superhorizon scale ($k_- \eta_- \ll 1$), the contributions of $s_0$ 
and $s_1$ are important, which we can calculate 
\begin{eqnarray}
&&s_0=
{\epsilon\ell^2 c_0\sqrt{\varepsilon_{1+}}\over 6\sqrt{1+c_0^2} 
\cosh(B_0\tau)}
\Biggl[-{\sin(2 k_- r_0) \over k_- \ell \sqrt{\epsilon_{1-}}}
\left(\sinh(B_0\tau)-c_0^{-1}\right)
\notag\\
&&+{\sqrt{1+c_0^2}\over c_0}\cosh(B_0\tau)
\left\{{\cos(2 k_- r_0)\over \sqrt{\epsilon_{1-}}}\left(1+{\eta_0 \over \ell}\right)-{1\over \sqrt{\epsilon_{1+}}}\left(4-{1\over c_0^2} +{2\eta_0 \over \ell}
\right)\right\}\notag\\
&&-\sinh(B_0\tau)\left({\cos(2 k_- r_0)\over \sqrt{\epsilon_{1-}}}
+{1\over \sqrt{\epsilon_{1+}}}\right)\left(1+{1\over c_0^2} +{\eta_0 \over \ell}\right)
+{\eta_0 \over c_0 \ell}\left({\cos(2 k_- r_0)\over \sqrt{\epsilon_{1-}}}
+{1\over \sqrt{\epsilon_{1+}}}\right)\Biggr]
\,,\label{eq:modulation-super}
\end{eqnarray}
and 
\begin{eqnarray}
&& s_1={\epsilon \ell^2 c_0 y_0\sqrt{\varepsilon_{1+}}\over 6 \sqrt{1+c_0^2} \cosh(B_0 \tau) }
\Bigg\{-{\sin(2 k_- r_0)\over k_- \ell \sqrt{\varepsilon_{1-}}}{y_0
\over c_0 \cosh^2(B_0 \tau)}+\Bigl({\cos(2 k_- r_0)\over \sqrt{\varepsilon_{1-}}}
+{1\over \sqrt{\varepsilon_{1+}}}\Bigr)\times \notag\\
&&\left[-{1\over \cosh^2(B_0 \tau)}\Bigl(1+{1\over c_0^2}+{\eta_0 \over \ell}\Bigr)
-{\eta_0 \over c_0y_0\ell }\Bigl(\cosh(B_0\tau)\sqrt{1+c_0^2}
-c_0 \sinh(B_0 \tau)+1+{\tanh(B_0 \tau) y_0
\over \cosh(B_0 \tau)}\Bigr)
\right]\notag\\
&&+{3\eta_0 \over c_0y_0\ell } {\cosh(B_0\tau) \sqrt{1+c_0^2}\over 
\sqrt{\varepsilon_{1+}}}\Biggr\}
\,,
\end{eqnarray}
where
\begin{eqnarray}
y_0=\sinh(B_0\tau)+c_0
\,.
\end{eqnarray}
The correction $s_1$ is derived from the time derivative terms of $\beta$ and $\til{\alpha}$, which plays as the order $\calO(1/\cosh(B_0 \tau))$ 
and then it can be ignored at the late time, although it can become a leading correction to (\ref{eq:modulation-super}) at the initial time. 

At the subhorizon scale ($k_-\eta_-\gsim  1$),  the dominant correction 
term is given by $s_2$, which is evaluated by  
\begin{eqnarray}
&& s_2=-{\epsilon \ell^2 k_- \eta_0  \sqrt{\varepsilon_{1+}}\over 6 }
\Bigg\{\Bigl({\cos(2 k_- r_0)\over \sqrt{\varepsilon_{1-}}}+{1\over \sqrt{\varepsilon_{1+}}}\Bigr){c_0\sinh(B_0 \tau)-1\over k_- \ell\sqrt{1+c_0^2}\cosh(B_0\tau)}-{2 k_-\eta_0
\over \sqrt{\varepsilon_{1+}}}\notag\\
&&+{c_0 \sin(2 k_- r_0)\over \sqrt{\varepsilon_{1-}}\sqrt{1+c_0^2} \cosh(B_0 \tau)}\left[{\sqrt{1+c_0^2}\over c_0} \cosh(B_0\tau)
\Bigl(1+{\eta_0\over \ell}\Bigr)-\sinh(B_0\tau)\Bigl(1+{1\over c_0^2}+{\eta_0\over \ell}
\Bigr)+{\eta_0\over c_0 \ell}\right]\Biggr\}
\,.\label{eq:s2}
\end{eqnarray}
\end{widetext}

\begin{figure}[h]
\begin{center}
\begin{tabular}{ccc}
\hspace{-1cm}
\includegraphics[width=7cm]{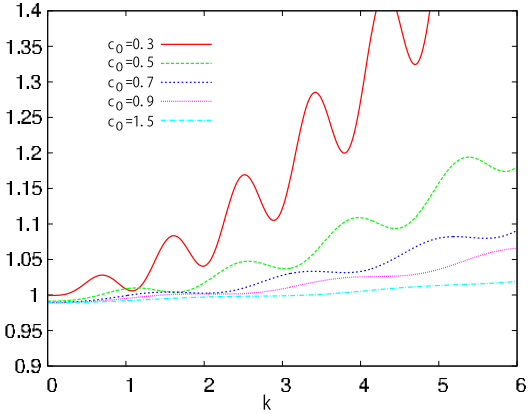} 
\hspace{-1cm}
\end{tabular}
\end{center}
%\vspace{-1cm}
\caption{We plot a modulation factor of power spectrum ($1+s_0+s_1+s_2$)
 given in (\ref{eq:def-dimpower}). We set $\epsilon=0.01\ell^{-2}$ and $\tau=\ell$ 
for each value of $c_0=\ell \sigma/4$ and assume slow-roll parameters on both sides of 
walls are equal as $\sqrt{\varepsilon_{1\pm}}=0.1$. 
 }
\label{fig:modulation-power}
\end{figure}

In order to discuss how those correction terms can be observed, 
we shall depict them numerically with 
 the specific values of parameters, which we 
choose 
\begin{equation}
\epsilon=10^{-2}\, \ell^{-2},~~{\rm and}~~\sigma\ge 0.4 \,\ell^{-1}
\,.
\label{eq:sigma-epsilon}
\end{equation}
This inequality guarantees to satisfy the condition $4 \epsilon<3\sigma^2$ given as 
(\ref{eq:small-epsilon}). In Fig.\ref{fig:modulation-power}, 
we show a modulation factor of power 
spectrum ($1+s_0+s_1+s_2$) 
 for different values of $\sigma$. 
It shows that the modulation factor 
can be larger in a subhorizon regime: $k_- \ell\,\gsim\, 1$. 
For a large value of $c_0$, 
the wall radius has still stayed to be constant. Namely $c_0>1$ gives 
a very tiny modulation, that is a 
nearly constant amplification in a power spectrum for $\tau=1$. 
Smaller value of $c_0$ shows that the modulation factor 
is increasing with oscillations in terms of $k$ 
(e.g. see the plot for $c_0=0.3$), which amplitude gets larger  
for a smaller value of $c_0$. 

\begin{figure}[h]
\begin{center}
\begin{tabular}{ccc}
\hspace{-1cm}
\includegraphics[width=6.5cm]{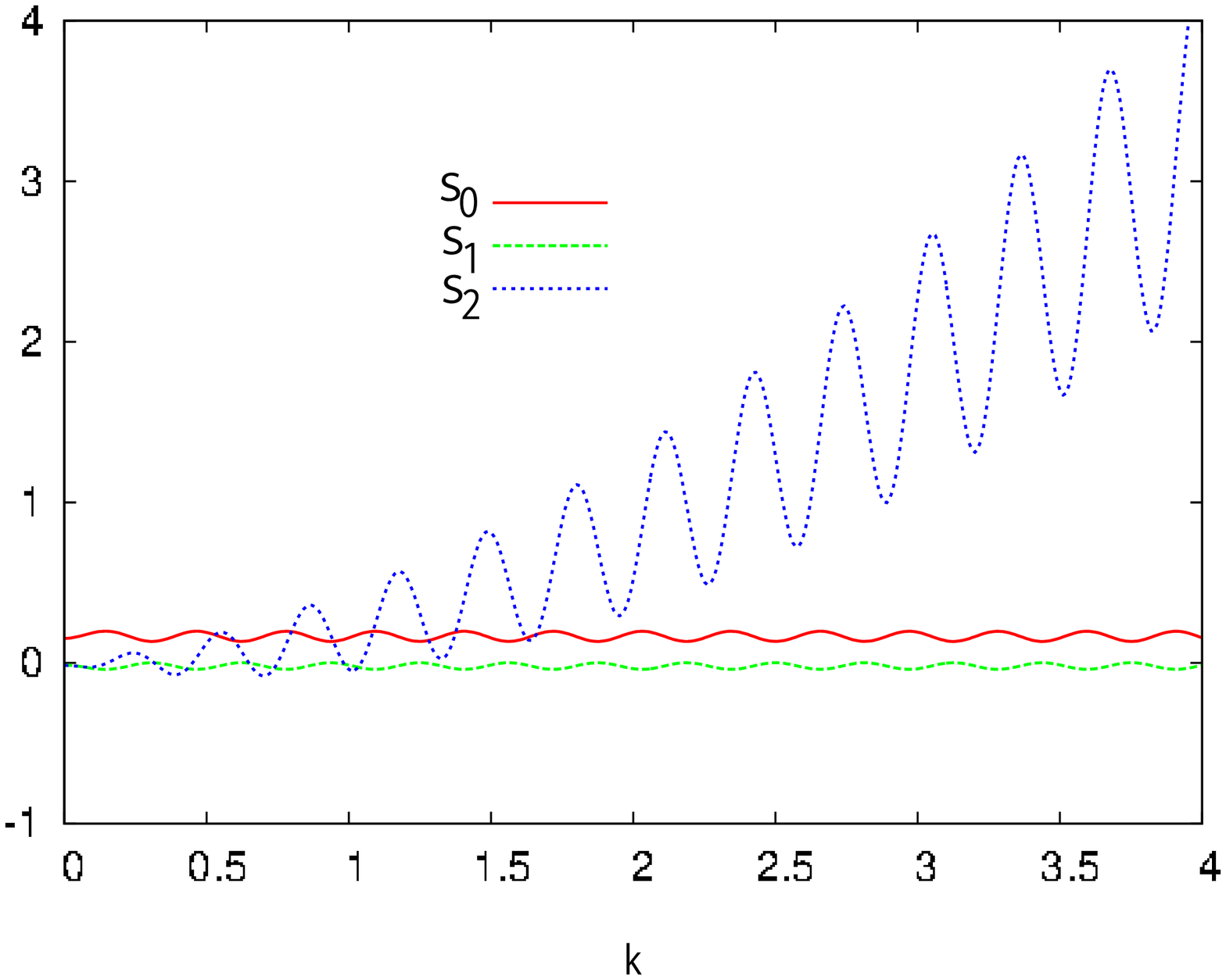} \\
\hskip -2cm
\includegraphics[width=6.5cm]{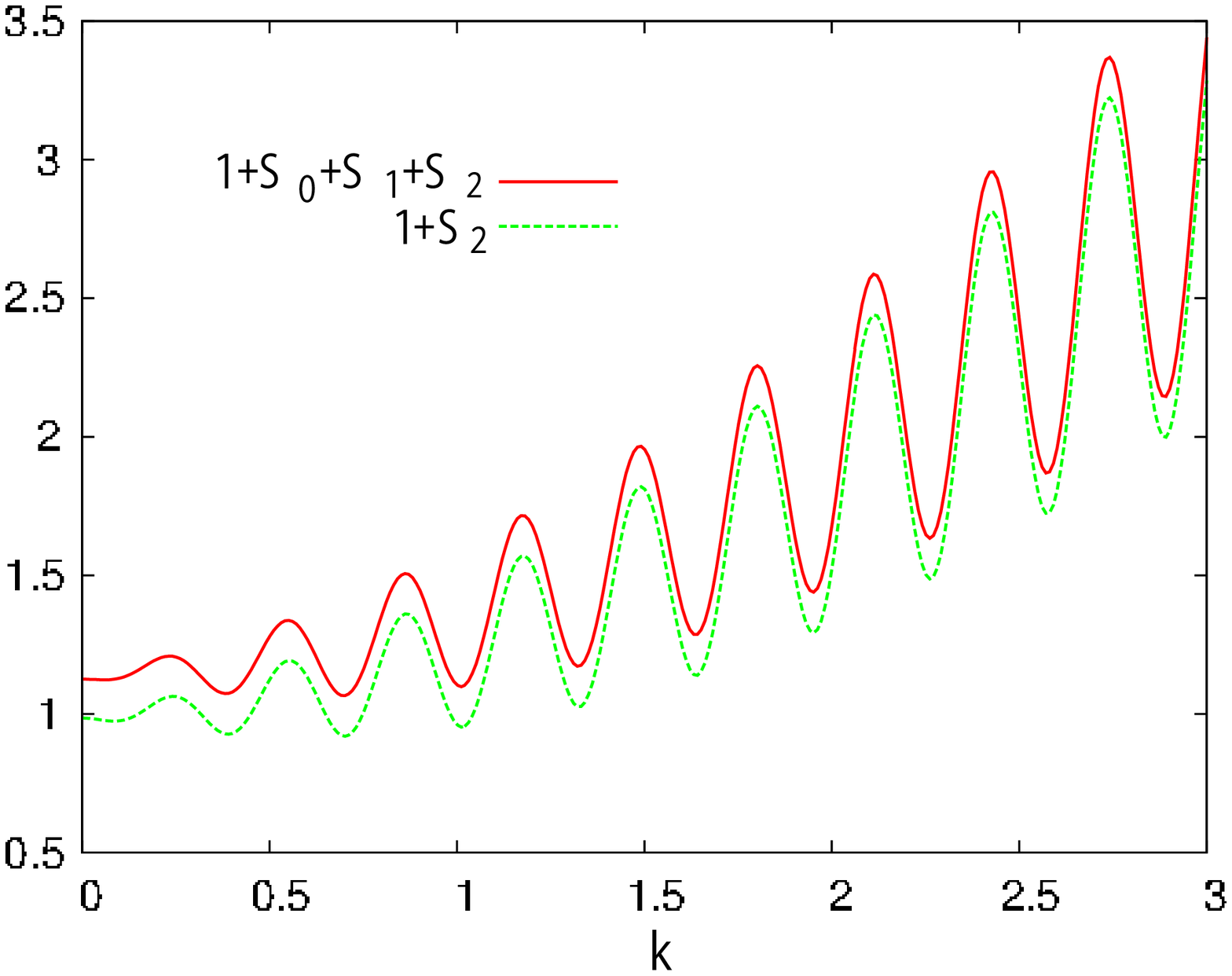}
\hspace{-1cm}
\end{tabular}
\end{center}
%\vspace{-1cm}
\caption{All parameters except for $c_0$ are same as used 
in  Fig.\ref{fig:modulation-power}. (Top)  We plot three contributions of $s_0, s_1$ and $s_2$ for $c_0=0.1$.
 (Bottom) We compare the term $1+s_2$ with total one.  
The term $s_2$ can give us a good approximation for describing 
the total modulation function. 
The increasing oscillation appears around $k_- \ell >1$.}
\label{fig:modulation-power2}
\end{figure}

In  Fig.\ref{fig:modulation-power2}, 
we plot
how each correction contributes to the total modulation of power spectrum. 
We find that the term $s_1$ can be ignored, 
while $s_0$ plays an important role in the superhorizon scale $k_-\ell \le 1$. 
On the other hand, 
$s_2$ plays an important role in the subhorizon scale, which gives an increasing 
function of $k$ with oscillations 
for $k_- \ell>1$. This subhorizon effect comes from the one 
proportional to $(k_- \eta_0)^2$ in (\ref{eq:s2}). That is transmitted wave propagating from the exterior universe.
 Note that  the contribution of $s_2$ becomes larger than unity, which means
 a breakdown of perturbative approach for $\epsilon$. 
The most effective term comes mainly from the one 
proportional to $(k_- \eta_0)^2$ in (\ref{eq:s2}). It gives the condition 
(\ref{eq:kmax}). 
We can then evaluate a breakdown 
scale $k_{\rm max}$ of our perturbative treatment for $\epsilon$ as 
\begin{equation}
{1\over 3}
\epsilon\ell^2 (k_-\eta_0)^2<\calO(1)\longrightarrow k_{\rm max}= 
\sqrt{{3\over \epsilon\ell^2} }{c_0 y_0(\tau)\over (1+c_0^2)\ell}
\,.
\label{eq:kmax}
\end{equation}
 For $c_0=0.1$, we find  a limited maximum value of wavenumber 
as $k_{\rm max}\simeq 2.2\ell^{-1}$, beyond which we cannot use this formula.

Of course, these plots also depend on 
 the parameter of the proper time  $\tau$ 
when the waves are scattered. 
However  the plot shows 
a similar behavior when $\tau<\ell $, 
although we set $\tau=\ell$ in Fig.\ref{fig:modulation-power}.  
For $c_0=0.1$, the parameter $\tau\,\gsim\, 4\ell $
 gives no modulation of power spectrum since the wall radius 
converges to a constant value. 

In the observational point of view, 
the observed power spectrum would be 
generated by a wave scattered at the wall around which an $e$-folding number takes about $50-60$. If an inflationary  period is much longer 
than $60~e$-foldings, it maybe difficult to observe such effect on the CMB power spectrum
because the observed modulation is generated at $\tau\gg \ell$ when
 the wall radius is almost constant. So 
we have to fine-tune the inflationary model whose period ends around $60~e$-foldings.
 In this case, the value of the modulation is determined by the 
scattered waves at the wall position 
when inflation has just started and the radius of the wall is expanding. 
It will show the deviation from a standard inflationary model.

Note that if the initial radius of the wall is much larger than the horizon scale $H^{-1}=\ell$, 
i.e., $c_0=\sigma\ell/4\ll 1$,
we may observe the original perturbation before the scattering at large scale.
 So there may not appear the modulation 
for such a larger scale perturbation.

%%%%%%%%%%%%%%%%%%%%%%%%%%%%%%%%%
\section{The effect of Different gravitational constants\label{effect-G}}
In the original Jordan frame, the gravitational constants are different 
in both vacua, which is given by the VEV's of 
the scalar field $\Phi$ by  Eq. (\ref{rel:G-Phi}).
Since the scalar field $\phi$ is given by Eq. (\ref{sol:phi-r-wallb}), 
the value changes from $\phi_-$ to $\phi_+$ suddenly 
near the bubble wall.
Hence the gravitational constants are almost constant 
both inside and outside of the bubble except near the wall region.
The VEV's in a false($+$) and true($-$) vacua are written by
\begin{equation}
\phi_+=b \mu,~~~\phi_-(=\phi_0)=-\mu
\,.
\end{equation}
Then we find
\begin{eqnarray}
\Phi_+=e^{(1+b)\mu /\sqrt{7}},~~~\Phi_-=1\,,
\end{eqnarray}
which gives 
\begin{eqnarray}
G_+=e^{-2(1+b)\mu /\sqrt{7}}G_N,~~~G_-=G_N\,.
\end{eqnarray}

\begin{figure}[h] 
\begin{center}
\includegraphics[width=6.5cm]{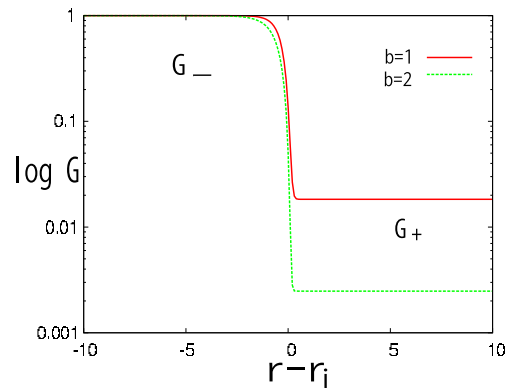} 
\end{center}
\caption{We plot the 
gravitational constants $G$ for $b=1$ and $b=2$, 
by setting $\mu=\sqrt{7}$ and $\lambda=1$.
We have used the radial coordinate $r$ 
in the Jordan frame.  $r_i$ denotes 
the  position of bubble wall. }
\label{fig:G-b120}
\end{figure}

Hence when we analyze the effect of the different gravitational 
constant in a false vacuum, we should   study 
the dependence of  $b$ on our results discussed in the previous section. 
If one wishes to set almost the same values for both gravitational constants, one needs to 
choose a negative value of $b$ such that $b\sim -1$, which implies that 
the value of $\phi$ in a false vacuum is close to that in the true vacuum ($\phi_+$). 
In Fig. \ref{fig:G-b120}, we show the gravitational constants for two values of $b$ ($b=1$
and $2$).

Since the parameter $b$ is related to  
  the surface density of the wall  $\sigma(b)$ as 
\begin{equation}
\sigma(b)={\sqrt{\lambda}\mu^3 \over 3 }(1+b)
\,,
\label{sol:sigma-b}
\end{equation}
which is one of the key parameters in our results, 
we can discuss the effect of the different gravitational constant 
by changing $\sigma$.  In fact, the ratio of two gravitational constants is given by
\begin{eqnarray}
{G_+\over G_-}=\exp\Big{[}-{12\sigma\over \sqrt{7\lambda}\mu^2}\Big{]}\,.
\end{eqnarray}

%%%%%%%%%%%%%%%%%%%%%%%%%
%\subsection{The effect of $\sigma(b)$}
Now we study how change of the gravitational constant outside of our universe affects 
on the final results obtained in the Einstein frame in the previous section. 
The effects are found mainly in two points; one is the bubble nucleation rate
and the other is the power spectrum of curvature perturbations, which we shall discuss 
in due order.  
\subsection{Nucleation rate of the bubble with 
 \\ different gravitational constant}
The bubble nucleation rate is given 
by (\ref{eq:Gamma-ImS}) and 
(\ref{eq:ImS}), which is described as 
\begin{equation}
\Gamma\simeq \exp\Bigl[-{\pi^2 \sigma\over 2H^3}\Bigr]\,,
\label{eq:Gamma-sigma}
\end{equation}
in the limit qof $\epsilon\rightarrow 0$.
This result shows that 
the larger value of $\sigma$ makes tunnelling process 
suppressed highly by a huge exponential factor. 
If $\sigma\simeq 0$, i.e.,  
$b\simeq -1$, which means the difference of gravitational constants
is very small, 
the nucleation probability 
becomes very high. 
This is easily understood as follows: 
By assuming that $\mu$ and $\lambda$ are fixed, 
when the value of $b$ changes from $-1$ to $\infty$,
 the surface tension $\sigma$ takes the value from $0$ to $\infty$.
It is because the wall barrier height is constant but the width becomes narrower 
as $b$ decreases.
When we change the gravitational constant in a false vacuum,
however, if we fix $\sigma$, i.e., the potential is 
given by the relation such that 
\begin{equation}
\lambda \mu^4 (\phi_+-\phi_-)^2\,,
\end{equation}
is constant,
the nucleation probability does not change.
Hence whether the universe with a different gravitational constant
is plausible or not depends on the choice of the potential form.

\subsection{Curvature perturbations}
The  power spectrum of curvature perturbations 
which reenter inside the horizon of our bubble universe
is evaluated by 
considering both reflected $\beta$ and transmitted $\tilde{\alpha}$ waves. 
The  modulation factor of a 
normalised power spectrum of curvature perturbations
(\ref{eq:def-dimpower})-(\ref{eq:s2}) 
is affected by changing $\sigma(b)$. 
In what follows, we assume $\mu=\sqrt{7}$ just for simplicity, and 
then  compare the case of $b=1$ with that of 
$b=2$.  If the surface density takes the value  $\sigma=0.4\ell^{-1}$ for $b=1$, 
we find $\sigma=0.6\ell^{-1}$ for $b=2$.
The gravitational constant $G_+$ for $b=2$ of the exterior 
of the bubble takes smaller value than the case of $b=1$ 
by the factor $e^{-2}\simeq 0.1$ 
because $(1+b)$ becomes three halves (see Fig.\ref{fig:G-b120}). 

\begin{figure}[h] 
\begin{center}
\includegraphics[width=7cm]{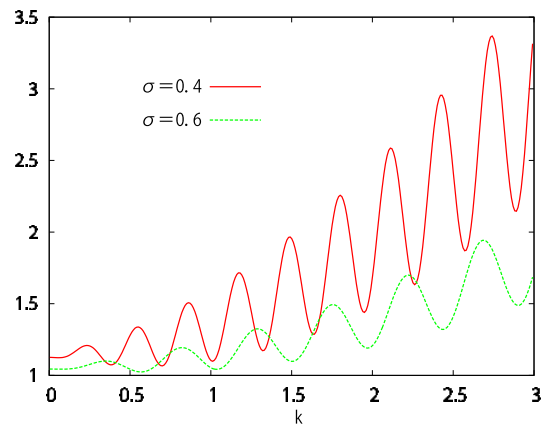} 
\end{center}
\caption{We plot the final power spectrums for $\sigma=0.4\ell^{-1}$ and 
$\sigma=0.6\ell^{-1}$.}
\label{fig:G-b12}
\end{figure}

For small value of $\epsilon$, which we assume through this paper,  
which have been obtained as (\ref{eq:modulation-super})-(\ref{eq:s2}), 
we plot the effect of difference of gravitational constants on the 
power spectrum  in Fig.\ref{fig:G-b12}. 
The modulation factor ($1+s_0+s_1+s_2$), which describes a correction from a standard 
scale-invariant spectrum, shows oscillatory bumps 
at a small scale $k_-\,\gsim\, 1/\ell$ and 
its amplitude increases for smaller $k_-$. It also increases for 
a smaller value of the parameter $\sigma$. 
By using (\ref{eq:s2}), the term such as $\cos(2 k_- r_0)$ 
gives the oscillation period $\Delta k\simeq \pi/r_0
\propto  \sigma$ for small $\sigma$. Therefore when $b$ becomes a half, 
the period $\Delta k$ becomes more sharp by twice. 

We conclude the effect of different gravitational constant on the 
final power spectrum as follows. The typical signature appears 
on a small scale and it gives an increasing oscillation. 
Especially, if 
the gravitational constants for both sides of the wall is close to each other, i.e.,
 $\sigma\approx 0$ ($b\approx -1$), the modulation of power spectrum becomes
  larger. The feature where an increasing oscillation appears on small scale is generated 
by the transmitted  $\tilde{\alpha}$ wave of perturbation in the 
exterior bubble universe (see (\ref{eq:til-alpha-e})), 
and can be constrained by the observation of CMB.

%%%%%%%%%%%%%%%%%%%%%%%%%%%%%%%%%%
%%%%%%%%%%%%%%%%%%%%%%%%%%%%%%
\section{Concluding remarks}

We have argued a possible scenario that each universe takes its own different value  
of gravitational constant in the context of the emergence of the bubble universes. 
In order to construct a model that makes different gravitational constants corresponding
 to several vacua, we adopt the 
JBD  theory, with which  
taking different values of the  VEV of the JBD scalar field, the universes with 
different  gravitational constants are achieved. In particular, we 
investigate the effect of different gravitational constant in the exterior universe on 
primordial curvature perturbations observed in the interior bubble universe. 

Via a conformal transformation, we can simply 
analyze the physics in the Einstein frame, that is 
the nucleation of bubbles, 
the evolution of bubble wall and matching curvature perturbations of the interior and 
exterior region. In the Jordan frame, it becomes clear that the final 
power spectrum of curvature perturbation is affected by the bubble 
boundary through the surface energy density 
$\sigma$ of the wall. 
The key parameter is $b$, which  is $b=\phi_+/|\phi_-|$, 
where $\phi_\pm$ is the VEV's of the JBD scalar field 
in the interior and exterior universes.
The parameter $b$ is related to 
the ratio of gravitational constants as $-2(b+1)\mu/\sqrt{7}=
\log( G_+/G_-)$. 
In Fig. \ref{fig:modulation-power}, for different values of $\sigma$, we show 
the modulation of the power spectrum, which  has several peaks with growing oscillation in  Fourier space. 
If the gravitational constant in the 
exterior universe gets 
 smaller, 
$b$ or $\sigma$ increases and then the modulation becomes smaller. 
Choosing a smaller parameter $\sigma$ will heighten the 
amplitude of peak shown in the modulation of power spectrum of curvature perturbations.

In our result, we assume that the $e$-foldings of  inflationary period is about $60$, 
because the initial modulation of perturbations will not be observed for  inflationary models
 with  the $e$-foldings more than $60$. 
In this paper, we have not discussed after inflation.
The result obtained from perturbations matching 
at later time such as a radiation dominated 
era will  be more interesting and important,
 although it is difficult to describe a trajectory 
of the bubble wall in an analytic form. 
In order to solve dynamics of wall in this phase, we may 
need numerical treatment. Along this direction, 
we hope to make a progress 
 in our future work. 

\acknowledgments
The work of YT is 
supported by a Grant-in-Aid through JSPS Fellow for Research Abroad H26-No.27. 
KM would like to thank DAMTP, the Centre for Theoretical Cosmology,
and Clare Hall in the University of Cambridge, where this work was started.
This work was supported in part by Grants-in-Aid from the 
Scientific Research Fund of the Japan Society for the Promotion of Science 
(No. 25400276). 

%%%%%%%%%%%%%%%%%%%%%%%%%%%%%%%%%
\appendix
\section{domain wall solution and surface energy for small $\epsilon$}
We consider the following potential
\begin{eqnarray}
V&=&V_0+\epsilon V_1
\nonumber \\
&=&{2\lambda\over (b+1)^4}(\phi-b\mu)^2(\phi+\mu)^2
\nonumber \\
&&
+{\epsilon\over (b+1)\mu}(\phi-b\mu)+\Lambda
\,.
\end{eqnarray}
There are three extrema
\begin{eqnarray}
\phi_+&=&b\mu-\epsilon{(b+1)\over 4\lambda\mu^3}\,,\\
\phi_{B}&=&{b-1\over 2}\mu+\epsilon{(b+1)\over 2\lambda\mu^3}\,,
\\
\phi_-&=&-\mu-\epsilon{(b+1)\over 4\lambda\mu^3}
\,.
\end{eqnarray}
$\phi_+$ and $\phi_-$ correspond to the local minima, while $\phi_B$ is 
the local maximum, at which the potential values are  by given
\begin{eqnarray}
V(\phi_+)&=&\Lambda+O(\epsilon^2)\,,
\\
V(\phi_{B})&=&\Lambda+{\lambda\over 8}\mu^4-{\epsilon\over 2}+O(\epsilon^2)\,,
\\
V(\phi_-)&=&\Lambda-\epsilon+O(\epsilon^2)
\,.
\end{eqnarray}
Although $\phi_+$ can be shifted by changing a free parameter $b$, 
 the potential barrier and the minimum values do not depend on $b$. 

Now we find a domain wall solution for this potential $V$.
Plugging $\phi=\phi_0(r)+\epsilon\phi_1(r)$ into the basic equation 
and expanding it  up to the first order of  small parameter $\epsilon$,
we find the following equations: 
\begin{eqnarray}
&&
\phi_0^{\prime\prime}+{1\over r}\phi_0^\prime-{dV_0\over d\phi}(\phi_0)=0\,,
\label{eq_phi0}
\\
&&
\phi_1^{\prime\prime}+{1\over r}\phi_1^\prime-{d^2V_0\over d\phi^2}(\phi_0)\phi_1
-{1\over (b+1)\mu}=0
\label{eq_phi1}
\,.
\end{eqnarray}
By using thin wall approximation, Eq. (\ref{eq_phi0}) gives the 0-th order 
domain wall solution as
\begin{eqnarray}
\phi_0={(b-1)\over 2}\mu+{(b+1)\over 2}\mu\tanh\left({r-r_i\over 2d}\right)\,.
\end{eqnarray}
Using this solution, we find
\begin{eqnarray}
{d^2V_0\over d\phi^2}(\phi_0)={4\lambda\mu^2\over (b+1)^2}\left[1-
{3\over 2\cosh^2\left({r-r_i\over 2d}\right)}\right]
\,.
\end{eqnarray}
If we ignore the deviation form unity near $r=r_i$ 
this term can be treated as a constant.
Then we find an approximate solution for $\phi_1$ as
\begin{eqnarray}
\phi_1=-{(b+1)\over 4\lambda\mu^3}\left[1-
{D\over d\cosh^2\left({r-r_i\over 2d}\right)}\right]
\,,
\end{eqnarray}
where $D$ is an arbitrary constant.

This solution $\phi=\phi_0+\epsilon \phi_1$ 
can be rewritten as
\begin{eqnarray}
\phi&=&{(b-1)\over 2}\mu-\epsilon{(b+1)\over 4\lambda\mu^3}
\nonumber \\
&&
+{(b+1)\over 2}\mu\tanh\left({r-r_i+\delta r_\epsilon\over 2d}\right)
\,,
\label{DW_sol_ep}
\end{eqnarray}
where $\delta r_\epsilon={\epsilon\over \lambda\mu^4}D$.
This  domain wall solution   is 
just shifted from one with $\epsilon=0$ by $-\epsilon{(b+1)\over 4\lambda\mu^3}$  
and $-\delta r_\epsilon$ in $\phi$- and $r$-directions, respectively.
The domain wall structure itself does not change.

Next we shall calculate the surface energy density,
which is defined by 
\begin{eqnarray}
\sigma&=&\int_0^\infty dr \left[
{1\over 2}(\partial_r\phi)^2+V_{\rm DW}(\phi)\right]
\label{DW_surface_energy}
\,,
\end{eqnarray}
where $V_{\rm DW}$ is the part of the potential which contributes to 
the structure of a domain wall.

Plugging the above solution (\ref{DW_sol_ep}) into the kinetic term and 
potential, we find up to the first order of $\epsilon$ as
\begin{eqnarray}
{1\over 2}(\partial_r\phi)^2&=&{\lambda\mu^4\over 8 \cosh^4
\left({r-r_i+\delta r_\epsilon\over 2d}\right)}\,,
\\
V(\phi)&\approx& {\lambda\mu^4\over 
8\cosh^4\left({r-r_i+\delta r_\epsilon\over 2d}\right)}
+{\epsilon\tanh\left({r-r_i\over 2d}\right)
\over 4\cosh^2\left({r-r_i\over 2d}\right)}
\nonumber \\
&+&
{\epsilon\over 2}
\left[\tanh\left({r-r_i\over 2d}\right)-1
\right]+\Lambda
\,.~~~~
\end{eqnarray}
The potential $V$ contains a back ground bulk energy, i.e., cosmological constants;
\begin{eqnarray}
V_{\rm bulk}=\Lambda-{\epsilon\over 2} \left[1-{\rm sgn}(r-r_i)\right]\,.
\end{eqnarray}
where 
 the sign function ${\rm sgn}(x)$
 is defined by
\begin{eqnarray}
{\rm sgn}(x)=\left\{\begin{array}{ccc}
1&{\rm if}&x>0\\
-1&{\rm if}&x<0\\
\end{array}
\right.
\,.
\end{eqnarray}
Since the bulk energy should not be included in the surface energy of the wall,
 we define
\begin{eqnarray}
V_{\rm DW}&:=&V(\phi)-V_{\rm bulk}
\nonumber \\
&=& {\lambda\mu^4\over 
8\cosh^4\left({r-r_i+\delta r_\epsilon\over 2d}\right)}
+{\epsilon\tanh\left({r-r_i\over 2d}\right)
\over 4\cosh^2\left({r-r_i\over 2d}\right)}
\nonumber \\
&&+{\epsilon\over 2}\left[\tanh\left({r-r_i\over 2d}\right)
-{\rm sgn}(r-r_i)
\right]\,.
\end{eqnarray}
In the thin-wall approximation ($d\ll r_i$), 
the surface energy $\sigma$ is evaluated approximately as
\begin{eqnarray}
\sigma&\approx&\int_{-\infty}^\infty dr \left[
{1\over 2}(\partial_r\phi)^2+V_{\rm DW}(\phi)\right]
\nonumber \\
&=& {\lambda\mu^4\over 4}\int_{-\infty}^\infty {dr \over  \cosh^4
\left({r-r_i+\delta r_\epsilon\over 2d}\right)}
\,,
\label{DW_surface_energy2}
\end{eqnarray}
which gives the same result as the case with $\epsilon=0$,
i.e.,
\begin{eqnarray}
\sigma\approx 
{\sqrt{\lambda}\mu^3(b+1)\over 3}
\,.
\end{eqnarray}
It is plausible because the domain wall structure does not change 
by the small $\epsilon$-modification of the potential.
%%%%%%%%%%%%%%%%%%%%%%%%%%%%%%%%


\begin{thebibliography}{99}
\bibitem{Susskind:2003kw} 
  L.~Susskind,
  %``The Anthropic landscape of string theory,''
  In {\it Universe or multiverse?} ed by B. Carr, pp247-266, 
hep-th/0302219. 
%%%%%%%%%%%%%%%%%%   bubble universe%%%%%%%%%%%%%%%%%%%%%%%%
\bibitem{Coleman:1980aw} 
  S.~R.~Coleman and F.~De Luccia,
  %``Gravitational Effects on and of Vacuum Decay,''
  Phys.\ Rev.\ D {\bf 21}, 3305 (1980). 
\bibitem{Sato:1980yn}
  K.~Sato,
  %``First Order Phase Transition Of A Vacuum And Expansion Of The Universe,''
  Mon.\ Not.\ Roy.\ Astron.\ Soc.\  {\bf 195} 467 (1981).
\bibitem{Sato:1981gv} 
  K.~Sato, H.~Kodama, M.~Sasaki and K.~Maeda, 
%``Multiproduction of Universes by First Order Phase Transition of a Vacuum,''
  Phys.\ Lett.\ B {\bf 108}, 103 (1982);
K. Maeda, K. Sato, M. Sasaki, H. Kodama 
%"Creation of De Sitter-schwarzschild Wormholes by a Cosmological First Order Phase Transition"
Phys.Lett. B {\bf 108},  98 (1982). 
\bibitem{Maeda:1985ye} 
  K.~Maeda,
  %``Bubble dynamics in the expanding universe,''
  Gen.\ Rel.\ Grav.\  {\bf 18}, 931 (1986).
\bibitem{Berezin:1987bc} 
  V.~A.~Berezin, V.~A.~Kuzmin and I.~I.~Tkachev,
  %``Dynamics of Bubbles in General Relativity,''
  Phys.\ Rev.\ D {\bf 36}, 2919 (1987).
\bibitem{Basu:1991ig} 
  R.~Basu, A.~H.~Guth and A.~Vilenkin,
  %``Quantum creation of topological defects during inflation,''
  Phys.\ Rev.\ D {\bf 44}, 340 (1991).
\bibitem{Bucher:1994gb} 
  M.~Bucher, A.~S.~Goldhaber and N.~Turok,
  %``An open universe from inflation,''
  Phys.\ Rev.\ D {\bf 52}, 3314 (1995), hep-ph/9411206.
\bibitem{Yamamoto:1995sw} 
  K.~Yamamoto, M.~Sasaki and T.~Tanaka,
  %``Large angle CMB anisotropy in an open universe 
%in the one bubble inflationary scenario,''
  Astrophys.\ J.\  {\bf 455}, 412 (1995), astro-ph/9501109. 
\bibitem{KeskiVakkuri:1996gn} 
  E.~Keski-Vakkuri and P.~Kraus,
  %``Tunneling in a time dependent setting,''
  Phys.\ Rev.\ D {\bf 54}, 7407 (1996), hep-th/9604151. 
\bibitem{Garriga:1998he} 
  J.~Garriga, X.~Montes, M.~Sasaki and T.~Tanaka,
  %``Spectrum of cosmological perturbations in the one bubble open universe,''
  Nucl.\ Phys.\ B {\bf 551}, 317 (1999), astro-ph/9811257. 
\bibitem{Aguirre:2007an} 
  A.~Aguirre, M.~C.~Johnson and A.~Shomer,
  %``Towards observable signatures of other bubble universes,''
  Phys.\ Rev.\ D {\bf 76}, 063509 (2007), arXiv:0704.3473.
\bibitem{Simon:2009nb} 
  D.~Simon, J.~Adamek, A.~Rakic and J.~C.~Niemeyer,
  %``Tunneling and propagation of vacuum bubbles on dynamical backgrounds,''
  JCAP {\bf 0911}, 008 (2009), arXiv:0908.2757.
\bibitem{Casadio:2011jt} 
  R.~Casadio and A.~Orlandi,
  %``Bubble dynamics: (nucleating) radiation inside dust,''
  Phys.\ Rev.\ D {\bf 84}, 024006 (2011), arXiv:1105.5497.
%%%%%%%%%%%%%%%%%%%%%%%%%%%%%%%%%%%
\bibitem{Weinberg:1988cp} 
  S.~Weinberg,
  %``The Cosmological Constant Problem,''
  Rev.\ Mod.\ Phys.\  {\bf 61}, 1 (1989). 
\bibitem{Brans:1961sx} 
P. Jordan, Zeil. Phys. {\bf 157}, 112 (1959);
  C.~Brans and R.~H.~Dicke,
  %``Mach's principle and a relativistic theory of gravitation,''
  Phys.\ Rev.\  {\bf 124}, 925 (1961).
\bibitem{Fujii:2003pa} 
  Y.~Fujii and K.~Maeda,
{\it The scalar-tensor theory of gravitation},
  (Cambridge. Univ. Press,  2003). 
\bibitem{extended_inflation} 
D. La and P. J. Steinhardt, Phya. Rev. Lett. {\bf 62}, 376 (1989); Phys. Lett. B {\bf 220},
 375 (1989);
D. La, P. J. Steinhardt, and E. Bertschinger, Phys. Lett. B {\bf 231}, 231 (1989). 
\bibitem{soft_inflation} 
A.L. Berkin, K. Maeda, J. Yokoyama,
%"Soft Inflation"
Phys. Rev. Lett. {\bf 65}, 141 (1990);
A.L. Berkin, K. Maeda,
%"Inflation in generalized Einstein theories" 
Phys.Rev. D{\bf 44}, 1691 (1991).
\bibitem{monopole_inflation} 
N. Sakai, J. Yokoyama, K. Maeda,
%"Monopole inflation in Brans-Dicke theory" 
Phys.Rev. D{\bf 59}, 103504 (1999),  gr-qc/9811024.
\bibitem{Guth:1980zm}
  A.~H.~Guth,
  %``The Inflationary Universe: A Possible Solution To The Horizon And Flatness
  %Problems,''
  Phys.\ Rev.\  D{\bf 23} 347 (1981).
\bibitem{Lyth-book}
D. H. Lyth and A. Liddle,
{\it Cosmological inflation and large-scale structure},
(Cambridge Univ. Press,  2009). 
\bibitem{Ade:2013ktc} 
  P.~A.~R.~Ade {\it et al.}  [Planck Collaboration],
  %``Planck 2013 results. I. Overview of products and scientific results,''
  Astron.\ Astrophys.\  {\bf 571}, A1 (2014), arXiv:1303.5062; 
  %\bibitem{Ade:2013uln} 
  P.~A.~R.~Ade {\it et al.}  [Planck Collaboration],
  %``Planck 2013 results. XXII. Constraints on inflation,''
  Astron.\ Astrophys.\  {\bf 571}, A22 (2014), arXiv:1303.5082.
\bibitem{Ertan:2007eq} 
  E.~Ertan and A.~Kaya, 
%``Boundary effects in local inflation and spectrum of density perturbations,''
  Gen.\ Rel.\ Grav.\  {\bf 40}, 1511 (2008), arXiv:0704.2284.
\bibitem{maeda1989}
%"Towards the Einstein-Hilbert Action via Conformal Transformation  "
K. Maeda, Phys.Rev. D{\bf 39} 3159 (1989).  

\bibitem{maeda1987}
%"Is the Compactified Vacuum Semiclassically Unstable? "
K. Maeda, Phys.Lett. B{\bf 186}, 33 (1987). 


%%%%%%%%%%%%%%%%%%%%%%%%%%%%%%%%%%%%%
%\bibitem{Allen:1985ux} 
%  B.~Allen,
  %``Vacuum States in de Sitter Space,''
 % Phys.\ Rev.\ D {\bf 32}, 3136 (1985).
\bibitem{Brandenberger:2012aj} 
%\bibitem{Martin:2000xs} 
  J.~Martin and R.~H.~Brandenberger,
  %``The TransPlanckian problem of inflationary cosmology,''
  Phys.\ Rev.\ D{\bf 63}, 123501 (2001), hep-th/0005209; 
R.~H.~Brandenberger and J.~Martin,
  %``Trans-Planckian Issues for Inflationary Cosmology,''
  Class.\ Quant.\ Grav.\  {\bf 30}, 113001 (2013), arXiv:1211.6753.
\bibitem{Danielsson:2002kx} 
  U.~H.~Danielsson,
  %``A Note on inflation and transPlanckian physics,''
  Phys.\ Rev.\ D{\bf 66}, 023511 (2002), hep-th/0203198.
\bibitem{Easther:2002xe} 
  R.~Easther, B.~R.~Greene, W.~H.~Kinney and G.~Shiu,
  %``A Generic estimate of transPlanckian modifications to the primordial power spectrum in inflation,''
  Phys.\ Rev.\ D{\bf 66}, 023518 (2002), hep-th/0204129.
\end{thebibliography}
\end{document}